\documentclass[]{article}
\usepackage{epsfig}
\title{{\em Ab initio} statistical mechanics of surface
adsorption \\ and desorption: I.~H$_2$O on MgO~(001) at low coverage}
\author{D.~Alf\`{e}$^{1,2,3}$ and M. J. Gillan$^{1,2}$ \\
$^1$London Centre for Nanotechnology, University College London \\ 
Gordon St, London WC1H 0AH, UK \smallskip \\
$^2$Department of Physics and Astronomy, University College London \\
Gower St, London WC1E 6BT, UK \smallskip \\
$^3$Department of Earth Sciences, University College London \\
Gower St, London WC1E 6BT, UK}

\begin{document}
\maketitle
\abstract{
We present a general computational scheme based on molecular
dynamics (m.d.) simulation for calculating the chemical potential
of adsorbed molecules in thermal equilibrium on the surface
of a material. The scheme is based on the calculation of
the mean force in m.d. simulations in which the height of
a chosen molecule above the surface is constrained, and
subsequent integration of the mean force to obtain the
potential of mean force and hence the chemical potential.
The scheme is valid at any coverage and temperature, so that
in principle it allows the calculation of the chemical
potential as a function of coverage and temperature. It
avoids all statistical mechanical approximations, except for the
use of classical statistical mechanics for the nuclei, and
assumes nothing in advance about the adsorption sites. From the
chemical potential, the absolute desorption rate of the molecules
can be computed, provided the equilibration rate on the surface
is faster than the desorption rate. We apply
the theory by {\em ab initio}
m.d. simulation to the case of H$_2$O on
MgO~(001) in the low-coverage limit, using the Perdew-Burke-Ernzerhof (PBE)
form of exchange-correlation. The calculations yield an {\em ab initio} 
value of the Polanyi-Wigner frequency prefactor, which is more than
two orders of magnitude greater than the value of $10^{13}$~s$^{-1}$
often assumed in the past. Provisional
comparison with experiment suggests that the PBE adsorption
energy may be too low, but the extension of the
calculations to higher coverages is needed before firm conclusions
can be drawn. The possibility of including quantum nuclear effects
by using path-integral simulations is noted.
}

\section{Introduction}
{\em Ab initio} modelling based on density functional theory (DFT)
has had a major impact on the understanding
of molecular adsorption and desorption at 
surfaces~\cite{reuter03,honkala05}. However, most of
this modelling has been of the static kind, in which structural
relaxation is used to calculate the energies of chosen
adsorbate geometries. For the interpretation of some
types of experiment, static calculations may suffice, but
in many practical situations entropy effects cannot be
ignored. This is clearly true when one considers surface phase
diagrams or adsorption isotherms, but it is also true for
rate processes such as thermal desorption. In a recent short
publication~\cite{alfe06}, we showed the possibility of using DFT
molecular dynamics simulation to calculate absolute desorption rates, with full
inclusion of all entropy effects. Our aims here are to describe the
general theory underlying that work, to outline its relevance to
the calculation of adsorption isotherms,
to present the simulations
themselves in more detail, and to indicate several 
developments that we plan to explore in future papers.

Although most DFT surface modelling has been static, the general idea
of {\em ab initio} statistical mechanics (AISM) applied to
condensed matter goes back well over 
10 years~\cite{smargiassi95,sugino95}. Over that period,
methods for calculating the {\em ab initio} free energy of
liquids and anharmonic solids have become 
well established~\cite{dewijs98,alfe99,gillan06}. In
surface science, one of the earliest publications describing a form
of AISM was that of Stampfl {\em et al.}~\cite{stampfl99} on the
O/Ru~(0001) system, in which they successfully calculated 
thermal desorption spectra, heat of adsorption and the surface
phase diagram. 

All previous AISM work that we know of on surface adsorbates
has been based on lattice-gas schemes,
which we specifically wish to avoid here. To make this point
clear, we recall that there are two fundamentally different approaches
to the statistical mechanics of adsorbates. In lattice gas theories,
it is assumed from the outset that the adsorbate 
atoms or molecules (for brevity, we shall simply say `molecules') can occupy
only specified surface sites. Calculations are then formulated in terms
of the occupancies of these sites and transition rates between them, the
model parameters being fixed either by appeal to experiment or (more recently)
to DFT calculations. A completely different approach is to regard
the adsorbate molecules as forming a two-dimensional fluid interacting
with the substrate. The methods we shall describe resemble the
second approach, and are also close to the AISM methods developed for
bulk liquids. A crucial point is that these bulk AISM methods
are designed to deliver free energies or chemical potentials
that are free of statistical-mechanical errors (more precisely,
these errors can be systematically reduced to any 
desired tolerance)~\cite{alfe01,alfe02}.
Here, we have the same aim for the AISM of adsorbates: the statistical
mechanical errors should be systematically controllable, so that
the only remaining error is due to the DFT exchange-correlation
approximation. In a lattice-gas approach, by contrast, 
the adoption of a discrete set of adsorption sites constitutes
an approximation that is difficult to eliminate, and this
is why we avoid this approach.

Coming to specifics, the modelling techniques that we wish to
develop should allow the calculation of the chemical potential
of a system of adsorbate molecules on the surface of a material
at any coverage and temperature, starting from a given DFT
exchange-correlation functional. We focus particularly on the
chemical potential, since this is the key to calculating all other
thermodynamic properties of the adsorbate, as well as the thermal
desorption rate measured in temperature programmed desorption (TPD)
experiments~\cite{masel96}. The techniques 
should not assume in advance anything
about the sites occupied by the adsorbate molecules, or about their
orientations, and should not rely on the
harmonic approximation for any of the degrees of freedom. Ultimately,
though not in this paper, we should like to be able to treat molecules
that dissociate and undergo chemical reactions. We want to be able
to include quantum nuclear effects, if necessary. Although we
assume for the moment that the {\em ab initio} total energy
function is provided by DFT, we recognise that DFT is often
inaccurate, and we want the techniques to be generalisable
to more accurate {\em ab initio} methods, such as 
quantum Monte Carlo~\cite{foulkes01}
or high-level quantum chemistry.

We present in the next Section a theoretical scheme which we believe
will be capable of allowing many of these aims to be accomplished. Our strategy
is to consider the system consisting of the adsorbate molecules on
the surface in complete thermal equilibrium with the gas phase.
All molecules in the system, including those in the substrate, are
treated on an equal footing. In this situation, there is a distribution
$\bar{\rho} ( z )$, giving the probability of finding adsorbate
molecules a distance $z$ from the surface. This distribution can
be expressed in terms of a potential of mean force (PMF):
$\bar{\rho} ( z ) = A \exp ( - \phi ( z ) / k_{\rm B} T )$, and the
PMF $\phi ( z )$ is the integral with respect to $z$ of the mean
force $\langle {\cal F}_z \rangle_z$ acting on the adsorbate molecules
when they are constrained to be at distance $z$ from the surface.
We shall show how these simple relationships allow the calculation
of the chemical potential of the adsorbate. We note that the same
relationships also yield the absolute rate of thermal desorption as
a function of coverage and $T$, provided all equilibration rates on the
surface are faster than the desorption rate. We shall show that, under
appropriate conditions, the theory yields the widely used
Polanyi-Wigner equation~\cite{polanyi32}, 
according to which the rate of change
of adsorbate coverage $\theta$ (molecules per adsorption site) in
a TPD experiment is given by:
\begin{equation}
d \theta / d t = - f \theta^n \exp ( - \Delta E / k_{\rm B} T ) \; ,
\label{eqn:polanyi_wigner}
\end{equation}
where $\Delta E$ and $n$ are the activation energy and reaction
order for desorption, and $f$ is a frequency prefactor. In the
analysis of experimental TPD data~\cite{redhead62,king75}, 
$f$ is commonly treated as
an empirical adjustable parameter, but the theory we present
is capable of yielding {\em ab initio} values for $f$.

The practical application
of the scheme reported in this paper (Sec.~\ref{sec:simulations}) 
consists of DFT simulations
of the H$_2$O molecule on the perfect MgO~(001) surface at several
temperatures in the limit of
zero coverage. We confine ourselves here to zero coverage, even though
the theory is equally valid at all coverages, simply because the
practical sampling requirements are easier to satisfy in this case.
The target we set ourselves is to calculate the chemical potential
of the adsorbate molecule in thermal equilibrium, with complete
inclusion of entropy effects, to an accuracy of better than
50~meV for the chosen exchange-correlation functional.
We shall see that even at low coverage entropy effects are very
important, and increase the frequency prefactor for thermal
desorption by over two orders of magnitude in the temperature range
of practical interest. 
In Sec.~\ref{sec:discussion}, we discuss the practical problems
involved in doing calculations at higher coverages, and possible
ways of overcoming them; we also discuss the generalisation to
include quantum nuclear effects, using DFT path-integral techniques.
Our conclusions are summarised in Sec.~\ref{sec:conclusions}

\section{Theory and techniques}
\label{sec:methods}

We begin our outline of theory and techniques by describing the
statistical mechanical relationships that allow us to calculate
the chemical potential of adsorbed molecules using PMF techniques.
We then summarise the arguments that allow the thermal desorption
rate to be deduced from this chemical potential in the case of
fast equilibration. At the end of this Section, we 
discuss the practical issues in calculating the PMF, and we
provide technical
details of the DFT molecular dynamics techniques used to perform
our simulations of the isolated H$_2$O molecule on the MgO~(001) surface.

\subsection{Calculation of the chemical potential using PMF}
\label{sec:chemical_potential}

The adsorbate molecules
are all identical to each other, and are called species A; the atoms in
the solid are called species B. We consider first the
situation where A-molecules in the gas phase are in complete
thermal equilibrium with those adsorbed on the surface. We assume that
A-molecules cannot penetrate into the bulk.
The situation we envisage is adsorption of molecules
on the surface of a perfect crystal, with this surface corresponding
to a particular crystal plane and being free of defects. This being
so, we can draw a ``reference plane'' parallel to the surface, whose
position is such that if translated outward by a few lattice
spacings along the surface normal it would be entirely in the gas
phase, and if translated inward by a similar amount it would be entirely
in the bulk. The origin of coordinates lies in this reference
plane, with the $z$-axis pointing along the outward normal.

When the entire system is in thermal equilibrium at a given temperature
$T$ and a given chemical potential of the A-molecules, there is
a density $\rho_0$ (number per unit volume) of A-molecules in the
gas phase, and a surface density $\sigma$ (number per unit area) 
on the surface.
In order to make $\sigma$ well defined, we have to decide what is
meant by an A-molecule being ``on the surface''. In order to do this, 
we define a ``monitor'' point for each A-molecule. If the molecule
is an atom, this monitor point is just the position of the nucleus. If it
is a molecule, it could be chosen as the position of a specified
nucleus, or as the the centre of mass, or in some other way. The particular
choice of monitor point will not (significantly) affect any of the
physical results of interest. To separate A-molecules in the gas from
those on the surface, we define a ``transition plane'', which is
parallel to the reference plane, but displaced from it by the 
$z$-displacement $z_0$. An A-molecule is counted as on the surface
if the $z$-coordinate of its monitor point is less than $z_0$,
and in the gas phase otherwise. To ensure that the value of 
$\sigma$ does not depend significantly on the position of the
transition plane, we choose $z_0$ so that the interaction of
an A-molecule with the solid is negligible for
$z \ge z_0$.

Since we have full thermodynamic equilibrium, the chemical potentials
$\mu_{\rm gas} ( \rho_0 , T )$ and $\mu_{\rm ads} ( \sigma , T )$
in the gas phase and on the surface must be equal. We are primarily
interested in situations where $\rho_0$ is so low that interactions
between A-molecules in the gas are negligible. It is then convenient
to express $\mu_{\rm gas}$ as:
\begin{equation}
\mu_{\rm gas} ( \rho_0 , T ) = k_{\rm B} T \ln ( \rho_0 \Lambda^3 ) +
\mu_{\rm gas}^\dagger ( T ) \; ,
\label{eqn:mu_gas}
\end{equation}
where the first term on the right is the chemical potential of
a perfect gas of structureless particles, whose mass $M_{\rm A}$ is
equal to that of the A-molecules, and $\Lambda$ is the thermal
wavelength $\Lambda = \left( h^2 / 2 \pi M_{\rm A} k_{\rm B} T \right)^{1/2}$.
If the A-molecules are atoms, then $\mu_{\rm gas}^\dagger ( T ) = 0$,
but if they are molecules, $\mu_{\rm gas}^\dagger ( T )$  
represents the $T$-dependent contribution to $\mu_{\rm gas} ( \rho_0 , T )$
from internal vibrations and rotations. In a similar way, it is 
convenient to express $\mu_{\rm ads} ( \sigma , T )$ as:
\begin{equation}
\mu_{\rm ads} ( \sigma , T ) = k_{\rm B} T \ln ( \sigma \Lambda^3 / d ) +
\mu_{\rm ads}^\dagger ( \sigma , T ) \; .
\label{eqn:mu_ads}
\end{equation}
where $d$ is an arbitrary fixed length. By analogy with the gas-phase
chemical potential, the first term on the right represents
the chemical potential of a gas of structureless particles of
mass $M_{\rm A}$ confined to the surface region. The ``excess term''
$\mu_{\rm ads}^\dagger ( \sigma, T )$ depends, of course on the value chosen
for $d$, but this is a convenient way to write $\mu_{\rm ads}$, because
in the limit where quantum nuclear effects can be ignored,
$\mu_{\rm ads}^\dagger ( \sigma, T )$ is then independent of
Planck's constant. This way of separating $\mu_{\rm ads}$ 
is particularly helpful
in the limit of low coverage, $\sigma \rightarrow 0$, when
$\mu_{\rm ads}^\dagger$ becomes independent of $\sigma$ and 
includes the adsorption energy of A-molecules, as well as the entropy
effects due to translations, vibrations and (hindered) rotations. For non-zero
$\sigma$, $\mu_{\rm ads}^\dagger$ also includes the energetic and
entropic effects of adsorbate-adsorbate interactions. With these
definitions, the combination of eqns~(\ref{eqn:mu_gas}) 
and (\ref{eqn:mu_ads}) with the
equilibrium condition $\mu_{\rm gas} = \mu_{\rm ads}$ gives:
\begin{equation}
\sigma / \rho_0 = d \exp \left[
\beta \Delta \mu^\dagger ( \sigma , T ) \right] \; ,
\label{eqn:sigma_over_rho}
\end{equation}
where 
$\Delta \mu^\dagger \equiv \mu_{\rm gas}^\dagger - \mu_{\rm ads}^\dagger$.
We refer to $\Delta \mu^\dagger$ as the ``excess chemical potential
difference'' (ECPD).
The adsorption isotherm ($\sigma$ as a function of $\rho_0$ or
gas pressure at constant $T$) can then be found 
by solving eqn~(\ref{eqn:sigma_over_rho})
for $\sigma$. In calculating the thermodynamic properties of the
adsorbate on the surface, it is therefore convenient to
focus on the quantity $\Delta \mu^\dagger ( \sigma , T )$.

We shall need later the relation between $\Delta \mu^\dagger$ and the
isosteric heat of adsorption $h_{\rm iso}$, which can
be defined as the negative slope of $\ln ( \rho_0 )$ plotted
against $1 / T$ at constant surface density $\sigma$.
From eqn~(\ref{eqn:sigma_over_rho}), we have:
\begin{equation}
h_{\rm iso} = - 
\left( \frac{\partial}{\partial \beta} 
\ln ( \rho_0 d / \sigma ) \right)_\sigma =
\left( \frac{\partial}{\partial \beta} 
\left( \beta \Delta \mu^\dagger \right) \right)_\sigma \; .
\end{equation}

To develop a strategy for calculating $\Delta \mu^\dagger$,
we consider the spatial distribution of A-molecules in
thermal equilibrium. Let $\rho ( {\bf r} ) \, d {\bf r}$ represent
the probability of finding the monitor point of any A-molecule
in volume element $d {\bf r}$ at position ${\bf r}$. If
${\bf r}_n$ is the dynamical variable representing the position of
the monitor point of A-molecule $n$ and there
are $\nu$ A-molecules in the system, then $\rho ( {\bf r} )$ is:
\begin{equation}
\rho ( {\bf r} ) = \left\langle \sum_{n = 1}^\nu 
\delta ( {\bf r} - {\bf r}_n ) \right\rangle \; ,
\end{equation}
where $\langle \, \cdot \, \rangle$ denotes thermal average. When
${\bf r} \equiv ( x, y, z )$ is near the surface, $\rho ( x, y, z )$
depends on $x$ and $y$ as well as $z$, and exhibits the
translational periodicity of the surface. We are not interested
here in the dependence on $x$ and $y$, so we consider the
distribution $\bar{\rho} ( z )$, which is the spatial average
of $\rho ( x, y, z )$ over $x$ and $y$. For $z$ far from the
surface, $\bar{\rho} ( z )$ is equal to the gas number density
$\rho_0$, but in the region of the surface, $\bar{\rho} ( z )$ will
have a large peak (or perhaps more than one peak), due to
A-molecules adsorbed on the surface. With the position 
$z_0$ of the ``transition plane'' chosen as above, this peak
is entirely in the region $z < z_0$, and for $z \ge z_0$,
$\bar{\rho} ( z )$ is very close to $\rho_0$. The surface
density $\sigma$ of adsorbed molecules is the integral of
$\bar{\rho} ( z )$ over the region $z < z_0$. It is convenient
to work with the distribution $y ( z ) \equiv \bar{\rho} ( z ) / \rho_0$,
which is normalised so that $\lim_{z \rightarrow \infty} y ( z ) = 1$.
Then we have:
\begin{equation}
\sigma / \rho_0 = \int_{- \infty}^{z_0} dz \, y ( z ) \; .
\label{eqn:sigma_over_rho_int_y}
\end{equation}
From this, we see that $y ( z )$ is closely related to
the ECPD $\Delta \mu^\dagger$. In fact,
from eqn~(\ref{eqn:sigma_over_rho}):
\begin{equation}
\exp \left[
\beta \Delta \mu^\dagger ( \sigma, T ) \right] =
\frac{1}{d} \int_{- \infty}^{z_0} dz \, y ( z ) \; .
\label{eqn:int_y}
\end{equation}

Standard m.d. simulation can be used to calculate the unnormalised
distribution $\bar{\rho} ( z ) \equiv \rho_0 y ( z )$ in the region of
$z$ where the adsorbed molecules spend most of their time. Accumulation
of a histogram for the probability distribution of $z$ suffices for
this purpose. However, in order to perform 
the integral in eqn~(\ref{eqn:int_y})
we need the {\em normalised} distribution $y ( z )$, and for this,
the entire region $z < z_0$ must be sampled. Under most
circumstances, simple accumulation of a histogram is not a
practicable way of doing this, because the probability of adsorbed
molecules sampling the region $z \sim z_0$ is so small. There
are several well known techniques for overcoming this ``rare-event''
problem. The technique employed in the present work uses the
potential of mean force (PMF)~\cite{sprik98}. In this approach, $y ( z )$ is 
expressed in terms of a PMF $\phi ( z )$, which plays the 
role of a $z$-dependent free energy:
\begin{equation}
y ( z ) = \exp \left[ - \beta \phi ( z ) \right] \; .
\label{eqn:PMF}
\end{equation}
By standard arguments, the
$z$-derivative $d \phi / d z$ is equal to minus the thermal
average of the $z$-component of the force acting on the
monitor point of a chosen A-molecule. Denoting by
$\langle {\cal F}_z \rangle_z$ the thermal average of the
$z$-component of the force acting on the molecule when the
$z$-component of its monitor point is constrained to have the
value $z$, we then have:
\begin{equation}
\phi ( z ) = \int_z^\infty d z^\prime \, \langle {\cal F}_z \rangle_{z^\prime}
\; ,
\label{eqn:MF}
\end{equation}
where $\langle {\cal F}_z \rangle_z$ is counted positive if directed outwards.
Combining eqns~(\ref{eqn:int_y}) and (\ref{eqn:PMF}), 
we obtain a formula for the
ECPD $\Delta \mu^\dagger ( \sigma , T )$ in
terms of the mean force:
\begin{equation}
\Delta \mu^\dagger ( \sigma, T ) = k_{\rm B} T \ln \left\{
\frac{1}{d} \int_{- \infty}^{z_0} dz \,
e^{- \beta \phi ( z )} \right\} =
k_{\rm B} T \ln \left\{
\frac{1}{d} \int_{- \infty}^{z_0} dz \, \exp \left[
- \beta \int_z^{\infty} d z^\prime \, \langle {\cal F}_z \rangle_{z^\prime}
\right] \right\} \; .
\label{eqn:ECPD_from_mean_force}
\end{equation}
The physical content of this expression is that
the ECPD is determined by the reversible
work (integral of mean force) performed on transporting the molecule
from one phase to the other. The calculation strategy is
to evaluate the mean force $\langle {\cal F}_z \rangle_z$
as a time average in a series of constrained DFT m.d. simulations
at a sequence of $z$ values, and to compute the mean force
integral numerically. This strategy can in principle
be applied at any temperature and coverage, provided ways
can be found of achieving adequate statistical accuracy.
It is one of the purposes of this paper to test the practical
feasibility of the strategy.

We note in passing an alternative way of calculating $y ( z )$,
which should also be effective, namely umbrella 
sampling~\cite{patey75,fichthorn02}.
In this, the probability distribution of the dynamical
variable being sampled is deliberately biased by adding to
the Hamiltonian a potential that acts on this variable.
In the present case, we would add a potential $V_{\rm up} ( z_1 )$
acting on the $z$-coordinate of the monitor point of a chosen
molecule. It is an exact result of
classical statistical mechanics that the probability 
distribution $\tilde{p} ( z )$ of the chosen molecule
is related to the distribution $y ( z )$ in the absence of
$V_{\rm up}$ by $\tilde{p} ( z ) = A y ( z ) 
\exp ( - \beta V_{\rm up} ( z ) )$, where $A$ is a constant.
If $V_{\rm up} ( z )$ is appropriately chosen (in particular,
if it is similar to $- \phi ( z )$), $\tilde{p} ( z )$ can
be fully sampled over the required region $z < z_0$ by
accumulation of a histogram, and $y ( z )$ can then be
recovered by multiplying by $\exp ( \beta V_{\rm up} ( z ) )$.

\subsection{Thermal desorption rate}
\label{sec:TD_rate}

The rate at which molecules desorb from the surface when there
is complete thermal equilibrium between gas and adsorbed
molecules can be derived by standard detailed-balance arguments,
which we recall briefly. (The arguments are well 
known~\cite{zangwill88}, but
we need to express them in a way that is consistent with
the present notation.) 
According to elementary statistical
mechanics, the outward flux $\kappa$ of molecules across the
transition plane is 
$\kappa = \rho_0 \left( k_{\rm B} T / 2 \pi M_{\rm A} \right)^{1/2}$,
where we make a negligible error in setting the number density at
the transition plane equal to $\rho_0$. If the sticking coefficient
$S$ of molecules arriving from the gas phase is unity, then the flux
$\kappa$ is entirely due to spontaneously desorbing molecules. But
if $S < 1$, then only the flux $S \kappa$ is due to spontaneous
desorption. We denote by $\gamma$ the probability per unit time
that a molecule spontaneously desorbs, so that
$\gamma \sigma = S \kappa$. It follows from
eqns~(\ref{eqn:sigma_over_rho}) and (\ref{eqn:sigma_over_rho_int_y}) that:
\begin{eqnarray}
\gamma & = &
\left( \rho_0 / \sigma \right) S \left( k_{\rm B} T /
2 \pi M_{\rm A} \right)^{1/2} =
S \left( k_{\rm B} T / 2 \pi M_{\rm A} \right)^{1/2} \left/
\int_{- \infty}^{z_0} dz \, y ( z ) \right. \nonumber \\
& = & S \left( k_{\rm B} T / 2 \pi M_{\rm A} \right)^{1/2}
\frac{1}{d} \exp \left[
- \beta \Delta \mu^\dagger \right]  \; .
\label{eqn:gamma}
\end{eqnarray}

Although eqn~(\ref{eqn:gamma}) is derived for conditions of complete thermal
equilibrium, it may sometimes be correct for the desorption rate
in a TPD experiment, even though this is an irreversible process.
A necessary condition for it to be correct is that the rate of
equilibration of the adsorbate on the surface be fast compared with the
desorption rate $\gamma$. In general, the desorption of a molecule
from the surface leaves behind a distribution of the remaining molecules
that is not typical of thermal equilibrium. If the molecules in the
region where the desorption happened do not have time to equilibrate
before the next desorption occurs in this region, then the use of
eqn~(\ref{eqn:gamma}) is no longer strictly correct. The equation can also fail
for another reason. In a TPD experiment, the molecules are deposited on
the surface at a very low temperature, and $T$ is steadily increased. If
the rate of temperature increase is so fast that the adsorbate system
has no time to equilibrate, then the desorption rate will be history
dependent, and eqn~(\ref{eqn:gamma}) will fail, even at low coverage. In the
simulations presented in Sec.~\ref{sec:simulations}, 
we shall study the relaxation
rates associated with diffusion and reorientation of the H$_2$O molecule
on MgO~(001), in order to test the correctness of the formula for
$\gamma$. 

\subsection{Practical calculation of PMF}
\label{sec:mean_force}

It is clear from the foregoing theory that the primary quantity
to be calculated is the mean force $\langle {\cal F}_z \rangle_z$,
from which we obtain the PMF and chemical potential by integration.
The calculation of $\langle {\cal F}_z \rangle_z$ by m.d.
simulation of H$_2$O on MgO~(001) raises a number of
practical issues. We recall first that we are free to
choose the molecular
``monitor position'' in different ways. In the present work,
we choose it to be the position of the water O atom.
At each time step of m.d., the electronic-structure calculation
yields a Hellmann-Feynman force on the core of the water O exerted
by the valence electrons and the other ionic cores. In constrained
m.d. in which the $z$-coordinate of water O is held fixed, 
${\cal F}_z$ is simply the $z$-component of this Hellmann-Feynman force.
However, we note a disadvantage of this choice of monitor point, 
which is that ${\cal F}_z$ does
not vanish even when the molecule is far from the surface, because
the water O oscillates with
the internal vibrational modes of the molecule. This means that
time averaging is needed to calculate $\langle {\cal F}_z \rangle_z$,
even when the constrained value of $z$ is far from the surface.
This problem can be avoided, and the statistical efficiency
somewhat improved, by choosing the monitor point to be the
centre of mass of the molecule, as we shall discuss in detail in
paper~II. 

Given our stated aim of calculating $\Delta \mu^\dagger$ correct to better
than 50~meV for a chosen exchange-correlation functional, we need
to give thought to the tolerance on the statistical error
in the calculation of $\langle {\cal F}_z \rangle_z$ at each $z$-value,
the number of $z$-values at which $\langle {\cal F}_z \rangle_z$ 
is needed, and the way in which the integrations of 
eqns~(\ref{eqn:MF}) and (\ref{eqn:ECPD_from_mean_force}) should be
performed. Tests show that in the absence of statistical errors
the integration of eqn~(\ref{eqn:MF}) can be performed
to obtain $\phi ( z )$ over the relevant $z$-range
to much better than the required tolerance if we have 
$\langle {\cal F}_z \rangle_z$ at $\sim 15$ roughly
equally spaced values, and if the integration is performed
by the trapezoidal rule. In the integral
$\int_{- \infty}^{z_0} dz \, \exp ( - \beta \phi ( z ) )$ needed
to obtain $\Delta \mu^\dagger$, the integrand varies
rapidly in the region of its maximum, but we obtain
the required accuracy by performing a cubic spline fit
to the the $\phi ( z )$ values. Coming to statistical
errors, we derive in Appendix~A a relationship between
the errors on the $\langle {\cal F}_z \rangle_z$ values
and the resulting statistical error on $\Delta \mu^\dagger$,
which allows us to estimate in advance the required duration
of the constrained m.d. simulations needed to reduce the error
below our required tolerance. 

%

The m.d. simulations were performed in the canonical 
$( N, V, T )$ ensemble, with a Nos\'{e} thermostat~\cite{nose84}. However,
careful attention needs to be give to the issue of
ergodicity, because the MgO substrate is close to being
harmonic at all temperatures of interest here, and the
internal H$_2$O vibrations are not only nearly harmonic
but also of much higher frequency than the lattice modes.
To ensure efficient transfer of energy between all degrees
of freedom, we add to the Nos\'{e} thermostat also an
Andersen thermostat~\cite{andersen80}, in which the velocities are periodically
randomised by drawing new velocities from the Maxwellian
distribution appropriate to the desired temperature.

In considering how to achieve the required statistical accuracy
on $\langle {\cal F}_z \rangle_z$ at each $z$, it is important
to note that, since $\langle {\cal F}_z \rangle_z$ is a static
thermal average, it is completely independent of the atomic
masses. This means that the masses used to generate the m.d.
trajectories do not have to be set equal to their physical
values, but can be chosen to improve the efficiency with
which phase space is explored. This is important for
the present system, because of the very high vibrational
frequencies of the H$_2$O molecule.
By artificially increasing the H mass, we can take 
a larger m.d. time step, without affecting the final
results. A convenient way to gauge the advantage
gained by altering the H masses is to consider the number of
m.d. time steps needed to achieve a specified
statistical accuracy in the calculation of
$\langle {\cal F}_z \rangle_z$ for a given value of $z$, the
statistical accuracy being given by the re-blocking analysis
described in Appendix A. In test calculations, we found that
if $m_{\rm H}$ is increased from 1 to 4, the time step
can be increased from 0.5 to 1.0~fs, and the number of
time steps needed to achieve the same accuracy is reduced
by a factor of two. Further increase of $m_{\rm H}$
makes little difference to the sampling efficiency. The
m.d. simulations reported here were performed with the
choice $m_{\rm H} = 8$, the masses of O and Mg atoms being
their physical values.

Our {\em ab initio} m.d. simulations were
performed with the projector-augmented-wave
(PAW) implementation of DFT~\cite{bloechl94,kresse99}, 
using the {\sc VASP} code~\cite{kresse96}. 
The plane-wave cut-off was 400~eV, and the augmentation-charge
cut-off was 605~eV.
We used core radii of 1.06~\AA\ for Mg and 0.80~\AA\ for O.
There have been many previous
DFT studies of H$_2$O on MgO~(001)~\cite{langel95}, 
and it is generally agreed that
the molecule does not dissociate at low coverage. The interaction of
the molecule with the surface is partly electrostatic, but there
is an important contribution from non-bonded interaction of the
O lone pair with surface Mg ions, and hydrogen bonding of H with
surface O. The choice of exchange-correlation functional for
such non-bonded interactions is not a trivial matter, and it can
make a large difference to interaction energies, as will be shown
in the next Section. Most of our calculations are performed with
the Perdew-Burke-Ernzerhof (PBE) functional~\cite{perdew96}, which is
usually taken to be one of the best parameter-free forms of
exchange-correlation functional. 
The simulations employ the usual slab geometry, and the simulation
conditions were decided on the basis of preliminary tests,
which are described next.

\section{Simulations: H$_2$O on MgO~(001) at low coverage}
\label{sec:simulations}

\subsection{Preliminary tests}
\label{sec:preliminary_tests}

Since our aim is to calculate the chemical potential and desorption
rate of an H$_2$O molecule in the limit of zero coverage
on the surface of semi-infinite bulk MgO, we have made
tests to ensure that the slab thickness, vacuum gap and size of
the surface unit cell are all large enough to bring the system
very close to this limit. The tests were done on the adsorption
energy $E_{\rm ads}$, defined in the usual way as the sum of the
energy $E ( {\rm H}_2 {\rm O} )$ of an isolated water molecule
and the energy $E ( {\rm MgO} )$ of the relaxed clean MgO slab minus
the energy $E ( {\rm H}_2 {\rm O} + {\rm MgO} )$ of the relaxed
system in which the molecule is adsorbed on the surface of the slab:
\begin{equation}
E_{\rm ads} = E ( {\rm H}_2 {\rm O} ) + E ( {\rm MgO} ) -
E ( {\rm H}_2 {\rm O} + {\rm MgO} ) \; .
\end{equation}
With this definition, $E_{\rm ads}$ is positive if the total
energy decreases when the molecule goes from the gas phase to
the adsorbed state. 

\begin{figure}
\centerline{a)
\includegraphics[width=2.5in,angle=-90]{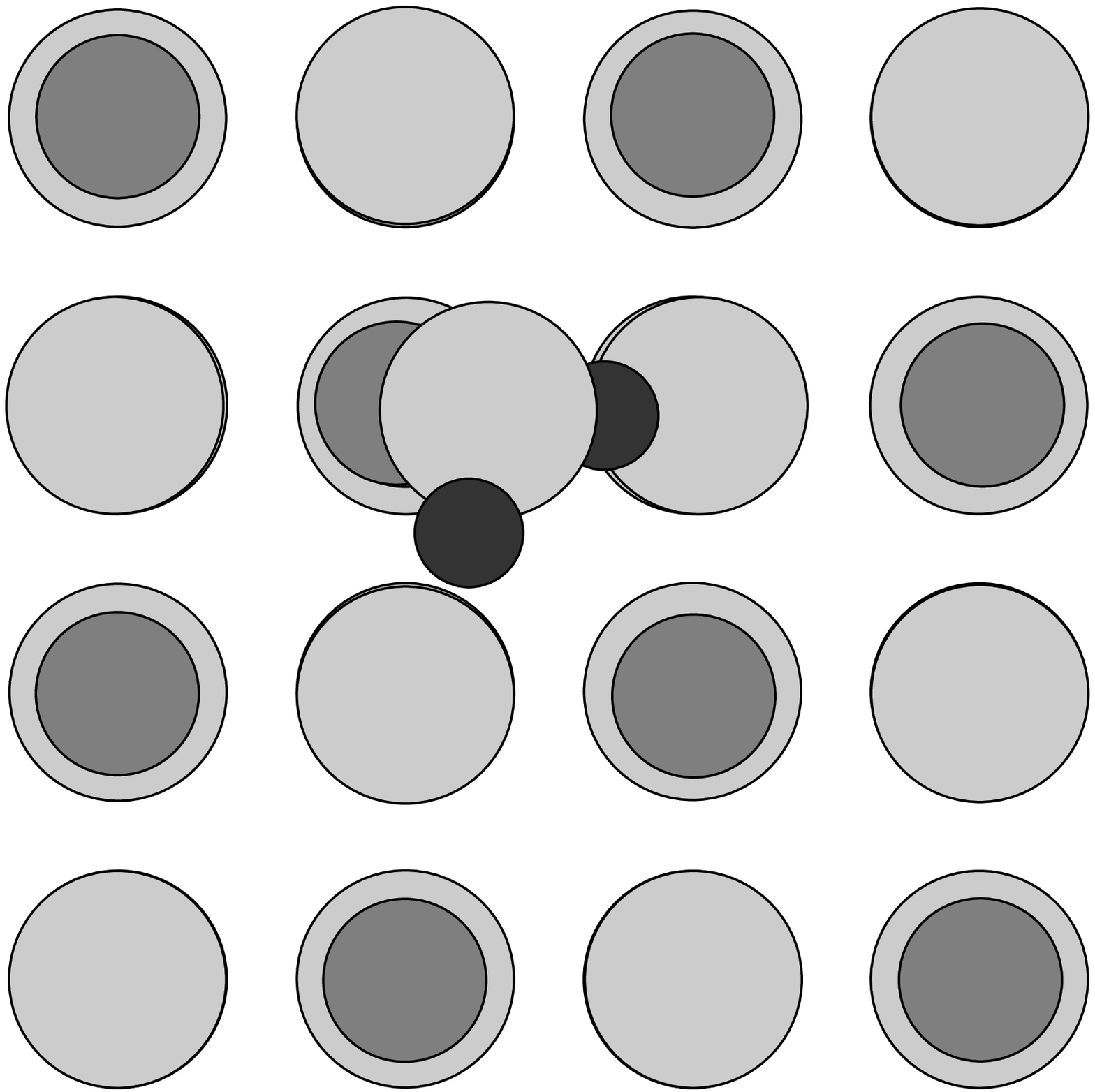} b)
\includegraphics[width=2.5in,angle=-90]{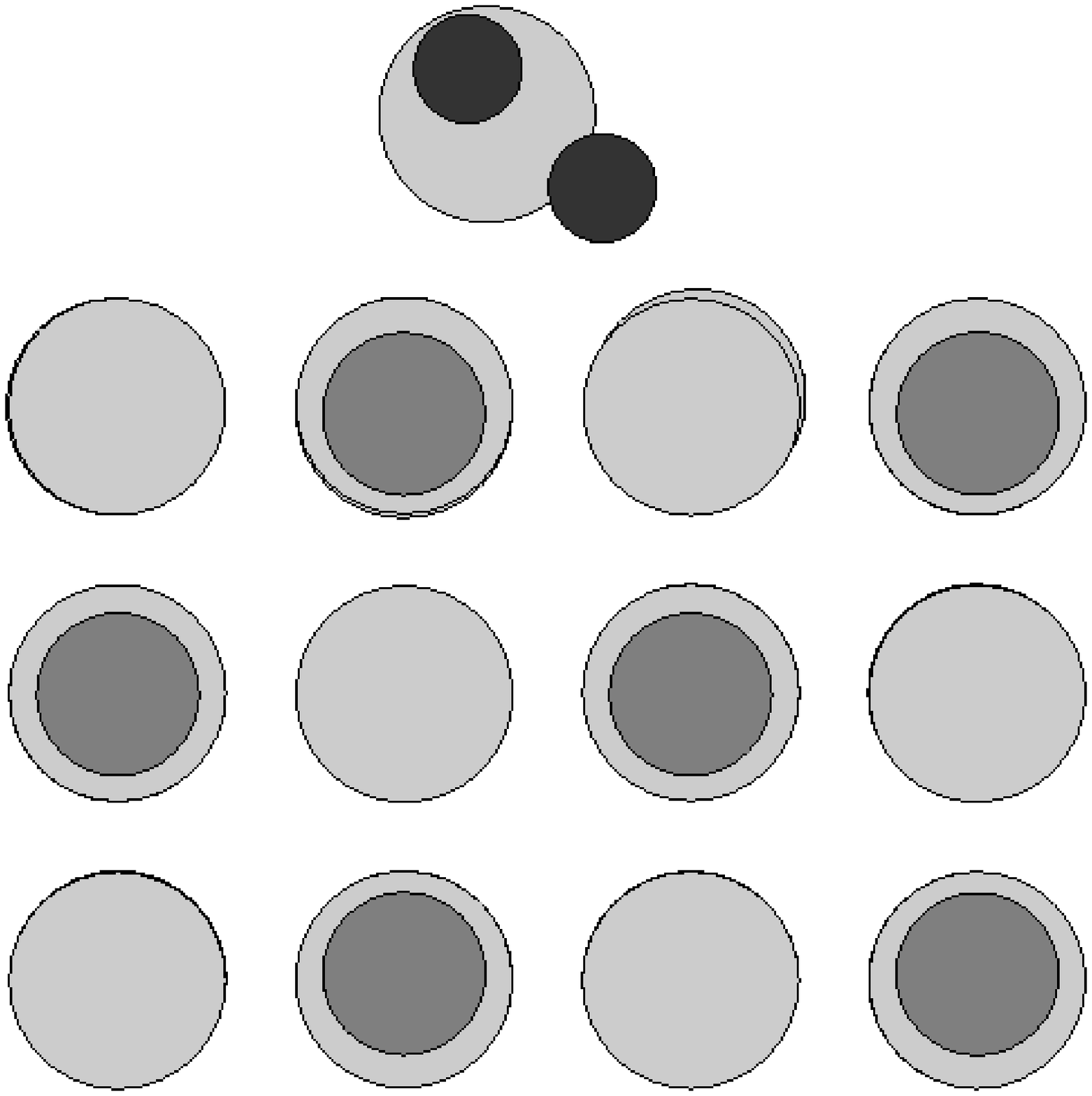}
}
\centerline{c)
\includegraphics[width=2.5in,angle=-90]{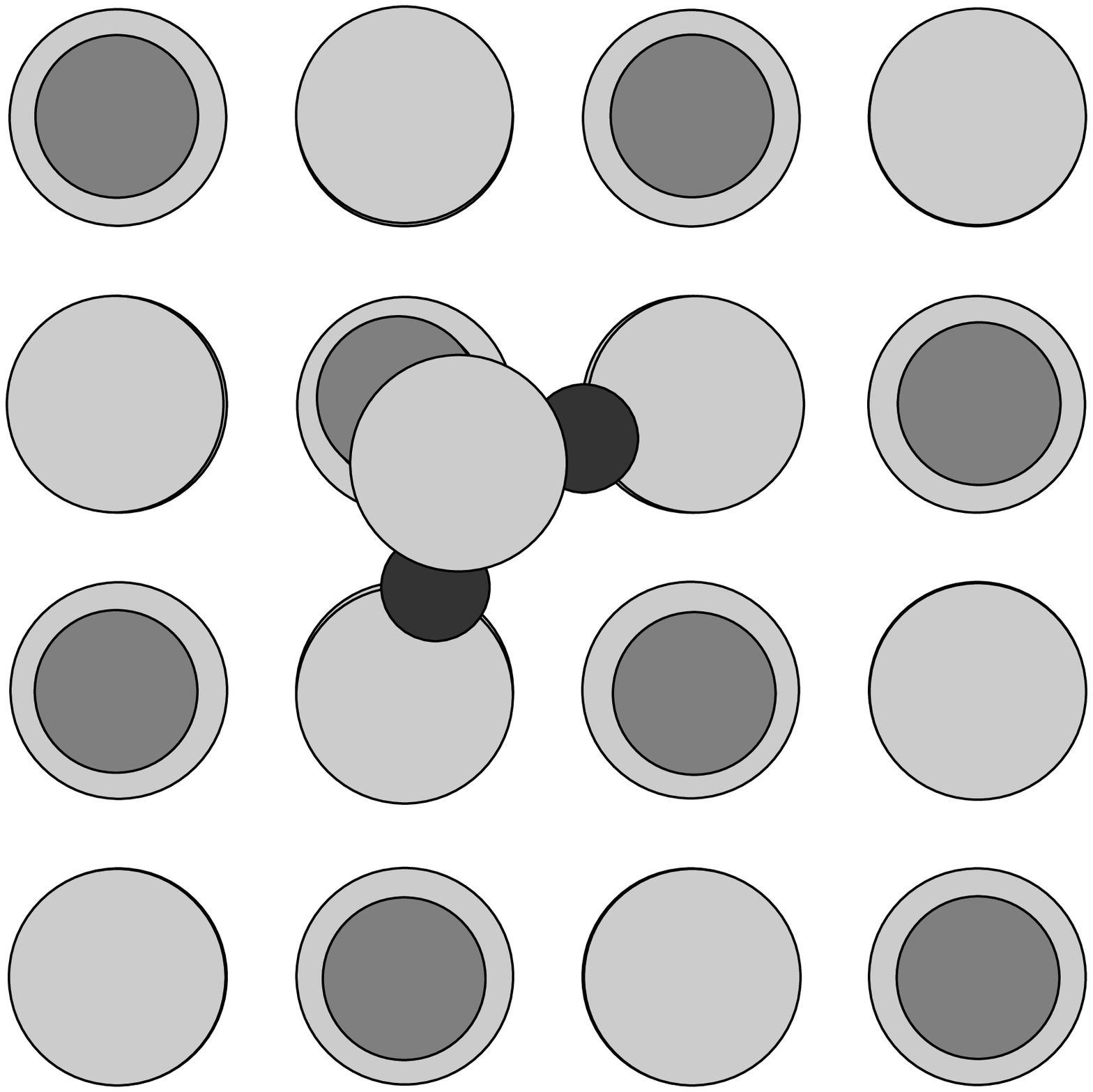} d)
\includegraphics[width=2.5in,angle=-90]{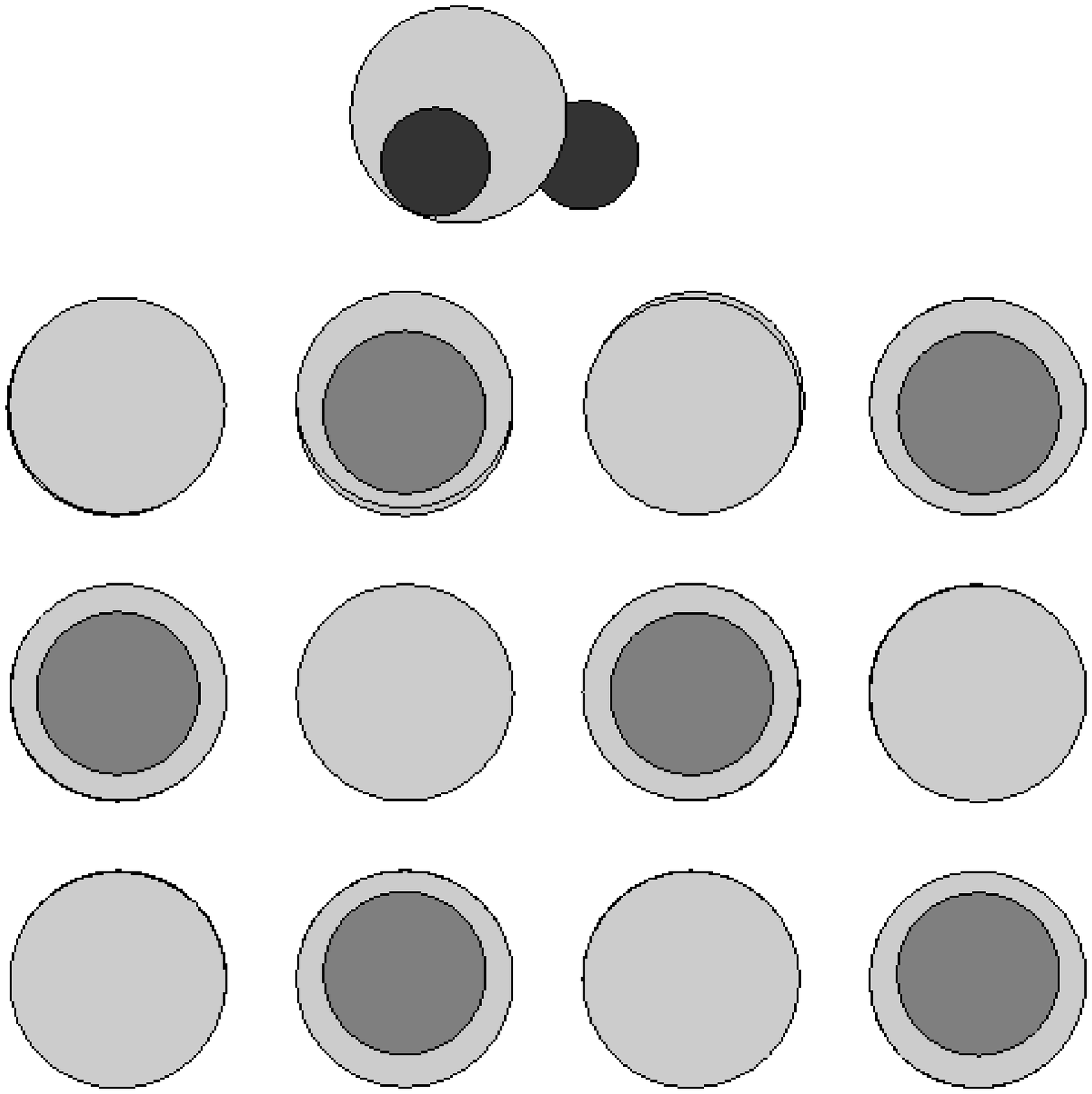}
}
\caption{Top and side views of the `tilted' ( panels a, b) and
`flat' (panels c, d) configurations of the H$_2$O molecule
adsorbed on the MgO~(001) surface. Oxygen:
large white spheres; Mg: medium grey spheres; H: small black spheres.}
\label{fig:adsorbed_config}
\end{figure}

With PBE calculations, the most stable adsorbed configuration
we have found (Fig.~\ref{fig:adsorbed_config}a,b) has 
the water O bound to a surface Mg,
one of the two O-H bonds directed to a surface O, and the other
tilted up at an angle to the surface plane. In this `tilted' geometry,
$E_{\rm ads}$ is converged to within $\sim 1$~meV with a slab
containing three layers of ions and vacuum width 
of 12.7~\AA\ (this is the distance between the layers of
surface ions that face each other across the vacuum gap). 
With a $2 \times 2$ surface unit
cell (16 ions per layer in the repeating cell), $E_{\rm ads}$ is
converged to better than 5~meV. With these settings, 
$\Gamma$-point sampling achieves convergence to within 5~meV.
The static adsorption energy given by PBE in this geometry is
0.46~eV. The calculations also reveal a second stable adsorbed
configuration having a slightly lower $E_{\rm ads}$ of 0.45~eV, in which
the water O is bound, as before, to a surface Mg ion, but with
the molecular plane almost flat on the surface, and with
both O-H bonds directed to surface O 
ions (Fig.~\ref{fig:adsorbed_config}c,d). When the
calculations are repeated with the LDA, we find 
that for the `flat' adsorbed geometry $E_{\rm ads}$ is 0.95~eV.
The large difference between the PBE and LDA values of $E_{\rm ads}$
indicates that quantitative agreement between theory and experiment
cannot be expected without calibrating DFT calculations against
more reliable methods. 

Guided by these tests, we have performed all the following calculations
with the 3-layer $2 \times 2$ slab (48 ions in the MgO slab per
repeating unit), a vacuum gap of 12.7~\AA, and $\Gamma$-point sampling.
Only the PBE approximation is used from now on. At all temperatures,
we set the lattice parameter equal to $4.23$~\AA, which is
the $T = 0$~K value in the bulk crystal, according to PBE.

\subsection{Results for PMF, chemical potential and desorption rate}
\label{sec:results}

We have stressed that the feasibility of the present scheme depends
on being able to make the statistical errors on 
$\langle {\cal F}_z \rangle_z$ small enough with simulation
runs of affordable length. 
We use the analysis presented in Appendix A to determine the
statistical errors on $\Delta \mu^\dagger$. We find that at $400$~K,
with typically 15 $z$-points and with runs of 40~ps duration at
each $z$-point, the statistical error on $\Delta \mu^\dagger$ is
less than 20~meV, which is much better than the accuracy we are
aiming for. The statistical errors are similar at other
temperatures.

\begin{figure}
\centerline{
\includegraphics[width=4.5in,angle=-90]{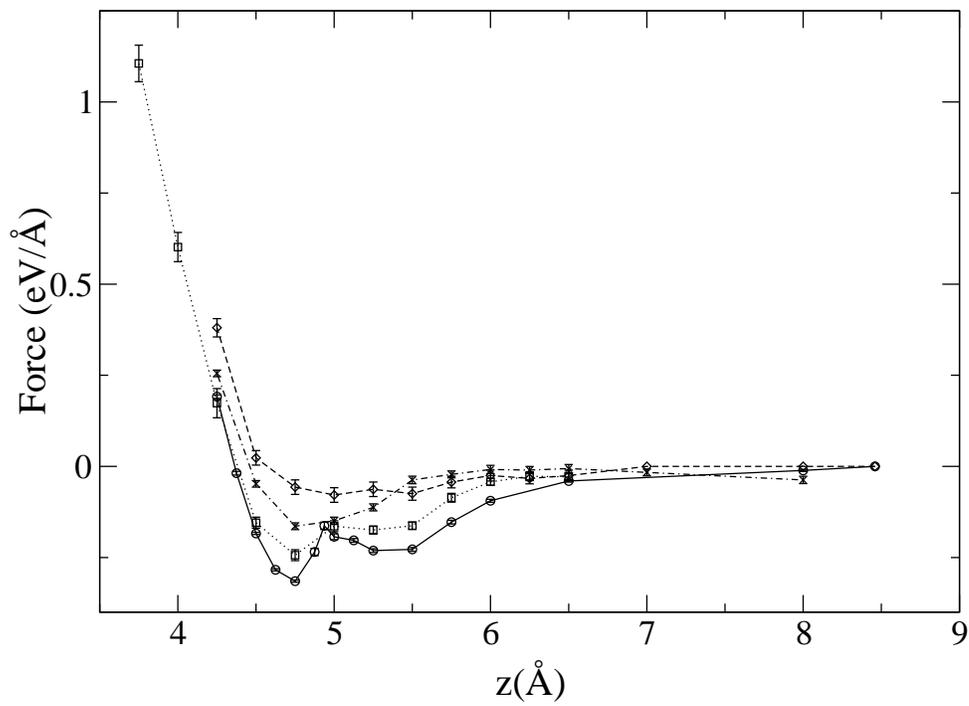}
}
\caption{The mean force $\langle {\cal F}_z \rangle_z$ on the
water O atom as function of its height $z$ above the surface
at $T = 100$, 300, 600 and 800~K (solid, dotted, chain and
dashed curves, respectively, are guides to the eye). Bars on
data points show statistical errors. Height $z$ is relative
to a fixed atom in the centre of the slab.}
\label{fig:MF}
\end{figure}

\begin{figure}
\centerline{
\includegraphics[width=4.5in,angle=-90]{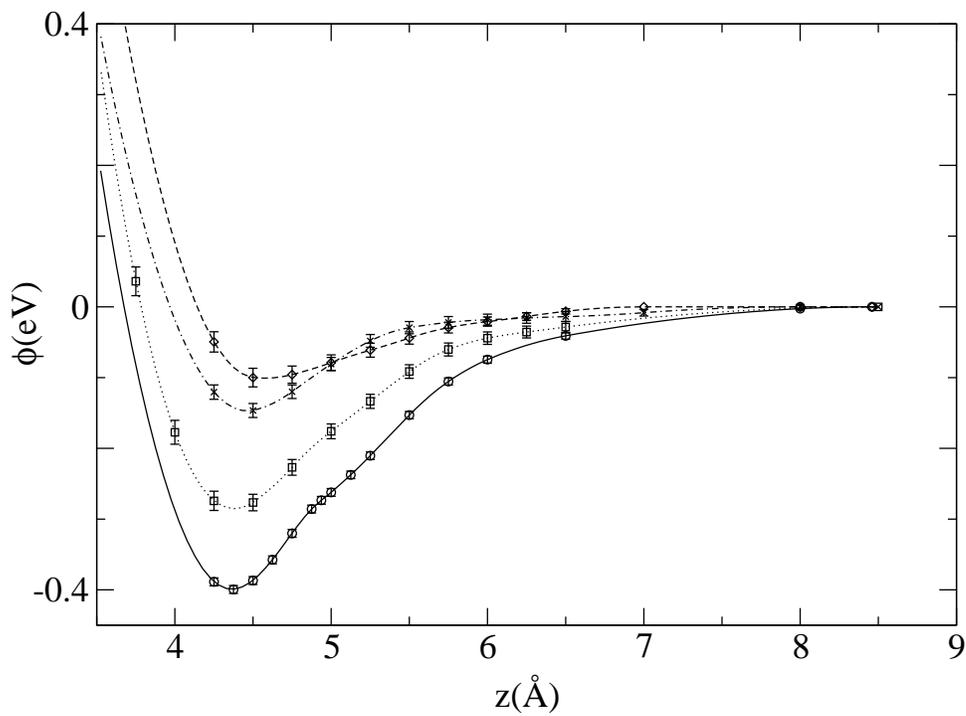}
}
\caption{Potential of mean force $\phi ( z )$ of
the water O atom as function of its height $z$ above the
surface at $T = 100$, 300, 600 and 800~K. Symbols and curves
have same meaning as in Fig.~\ref{fig:MF}.}
\label{fig:PMF}
\end{figure}

The mean force and the PMF from our simulations at four
temperatures are reported in Figs.~\ref{fig:MF}
and \ref{fig:PMF}. We note the strong
temperature dependence of both quantities. Interestingly,
$\langle {\cal F}_z \rangle_z$ shows a double minimum at low $T$,
which we believe is due to the fact that the
characteristic orientation of the H$_2$O molecule
changes substantially with $z$. The temperature dependence
of the well-depth in $\phi ( z )$ is closely
related to the entropy part of $\Delta \mu^\dagger$, as can be
seen by representing $\phi ( z)$ in the simple form 
$\phi ( z ) = \phi_{\rm min} + \frac{1}{2} c ( z - z_{\rm min} )^2$, where
$\phi_{\rm min}$ is the (negative) value of $\phi ( z )$ and $c$ is
the curvature at the bottom of the well. With this approximation,
we have:
\begin{equation}
\Delta \mu^\dagger = k_{\rm B} T \ln \left[
d^{-1} ( 2 \pi k_{\rm B} T / c )^{1/2}
\exp ( - \beta \phi_{\rm min} ) \right] =
- \phi_{\rm min} ( T ) +
\frac{1}{2} k_{\rm B} T \ln ( 2 \pi k_{\rm B} T / c d^2 ) \; .
\label{eqn:approx_delta_mu}
\end{equation}
Then the entropy $s = - \partial \Delta \mu^\dagger / \partial T$ is:
\begin{equation}
s = - \frac{1}{2} k_{\rm B} \left(
1 + \ln ( 2 \pi k_{\rm B} T / c d^2 ) \right) +
\partial \phi_{\rm min} / \partial T \; .
\end{equation}
The first term on the right represents the entropy contribution
from confinement of the $z$-coordinate of the molecule. The second
term, which is positive (the well-depth decreases with increasing
$T$), is due to the confinement of its translational and rotational
degrees of freedom, as will be discussed in more detail below.
Our calculated values of $\Delta \mu^\dagger$ at the four
simulation temperatures vary rather linearly with
$T$, and this implies that the isosteric heat of adsorption
$h_{\rm iso}$ deviates only slightly from the $T = 0$~K value
of the adsorption energy $E_{\rm ads}$.

\begin{figure}
\centerline{
\includegraphics[width=4.5in,angle=-90]{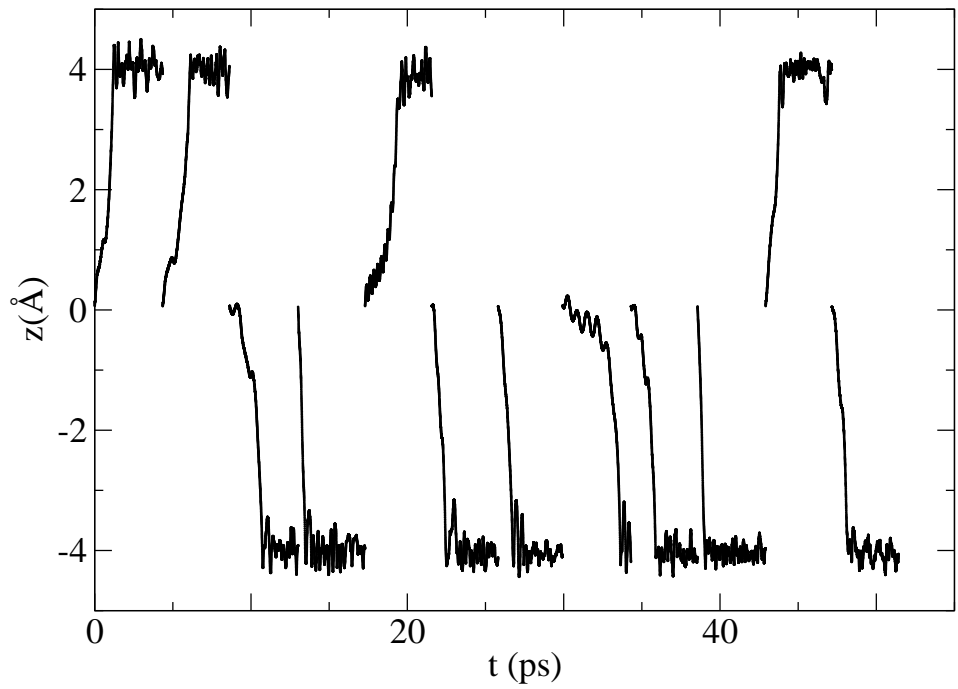}
}
\caption{Set of 12 trajectories from simulations used to
determine sticking coefficient $S$. Plots show $z$-coordinate
(\AA\ units) of water O atom relative to centre of vacuum gap between
slabs. 
}
\label{fig:sticking}
\end{figure}

In order to obtain the temperature-dependent desorption rate $\gamma$,
we need to know the sticking coefficient $S$. We expect this
to be close to unity, because the PMF shows that there
is no barrier to adsorption. To test whether $S$ deviates
significantly from zero, we have conducted a series of
simulations in which the 3-layer MgO slab has already been
well equilibrated at a given temperature $T$; the H$_2$O molecule
is initially in the middle of the vacuum gap, and its atoms are given
random velocities corresponding to that $T$. The subsequent
time evolution is then monitored. We show in Fig.~\ref{fig:sticking} the
results of 12 such simulations at $T = 400$~K. We see that
in all cases H$_2$O becomes adsorbed on the surface, and there
is no indication of desorption in the period of several ps after
adsorption, during which time the molecule will have become
well equilibrated on the surface. In the light of this,
it is an accurate approximation to set $S = 1$.

\begin{figure}
\centerline{
\includegraphics[width=4.5in,angle=-90]{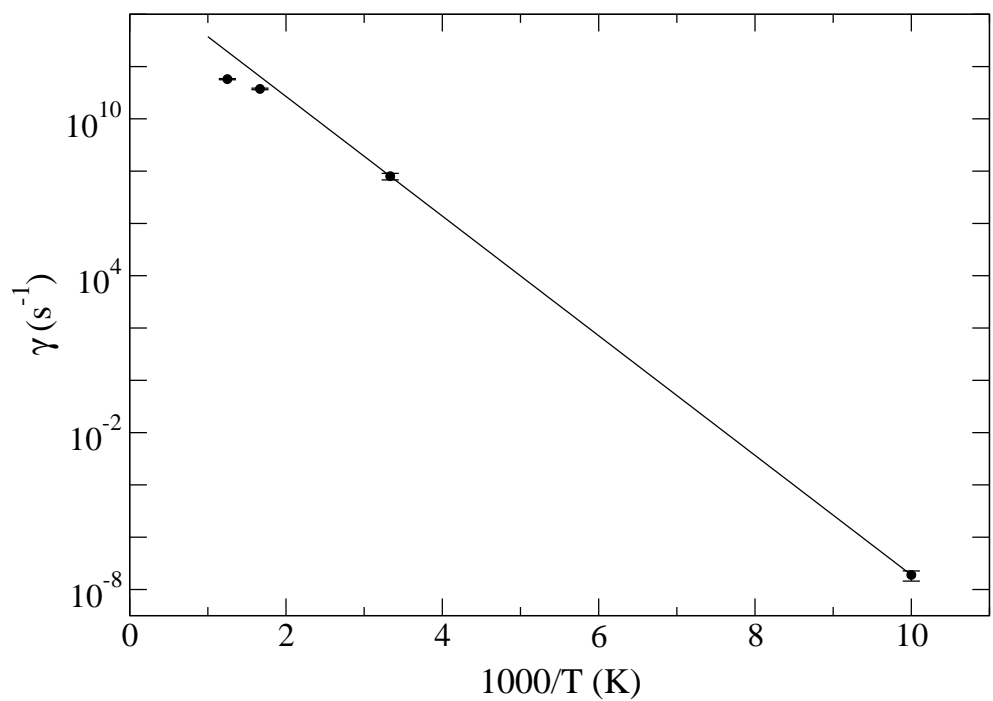}
}
\caption{Arrhenius plot of the desorption rate $\gamma$ of H$_2$O
from MgO~(001) calculated using the PBE exchange-correlation
functional. Bars on calculated values show statistical errors.
The straight line is drawn to pass through the calculated
values at the two lowest temperatures.}
\label{fig:arrhenius_gamma}
\end{figure}

Numerical values of the desorption rate $\gamma$ as a 
function of $T$ calculated
from eqn~(\ref{eqn:gamma}) are shown as an Arrhenius plot in 
Fig.~\ref{fig:arrhenius_gamma}. We 
have only a few data points, but the Arrhenius plot 
appears to rather straight, except at 
the highest $T$. This means that our results are 
fairly well described by the Polanyi-Wigner 
formula, eqn~(\ref{eqn:polanyi_wigner}). We note that for the 
present case of low-coverage 
desorption in the absence of dissociation, the reaction 
order $n$ is unity, so that the 
Polanyi-Wigner formula becomes 
$\gamma = - \theta^{-1} d \theta / d t =
f \exp ( - \Delta E / k_{\rm B} T )$. 
If we pass a straight line through the 
points at the two lowest temperatures in
fig.~\ref{fig:arrhenius_gamma}, we obtain an 
activation energy $\Delta E = 0.454$~eV, and
a prefactor $f = 2.7 \times 10^{15}$~s$^{-1}$.
We note that the activation energy is almost
exactly equal to the zero-temperature
static adsorption energy $E_{\rm ads} = 0.46$~eV.
In the temperature range 
$200 - 300$~K, where thermal desorption 
of H$_2$O from MgO~(001) is 
experimentally observed at sub-monolayer coverage, 
the calculated prefactor of 
$2.7 \times 10^{15}$~s$^{-1}$ is thus enhanced by a factor 
of $\sim 270$ above the value of 
$10^{13}$~s$^{-1}$ often used in analysing TPD 
measurements on this system in 
the past~\cite{stirniman96,xu96}. 

The enhancement of the prefactor arises from 
the $T$-dependence of the PMF. On 
substituting the approximate form of 
$\Delta \mu^\dagger$ given by eqn~(\ref{eqn:approx_delta_mu}) into 
eqn~(\ref{eqn:gamma}), we find:
\begin{equation}
\gamma = S \left( c \left/ 4 \pi^2 M_{\rm A} \right. \right)^{1/2} 
\exp ( \beta \phi_{\rm min} ) \; .
\end{equation}
The factor $( c / 4 \pi^2 M_{\rm A} )^{1/2}$ is the 
oscillation frequency of the centre 
of mass of the molecule along the surface normal. 
If $\phi_{\rm min}$ varies 
roughly linearly with $T$, then 
$\phi_{\rm min} = - E_{\rm ads} + 
T ( \partial \phi_{\rm min} / \partial T )$, so 
that $\gamma = S ( c / 4 \pi^2 M_{\rm A} )^{1/2} 
\exp ( k_{\rm B}^{-1} \partial \phi_{\rm min} / \partial T ) 
\exp ( - \beta E_{\rm ads} )$. 
The enhancement factor is thus $\exp ( k_{\rm B}^{-1} \partial 
\phi_{\rm min} / \partial T )$, and it comes from the 
part of the adsorption entropy 
associated with confinement of the 
translational and rotational degrees of freedom 
other than oscillation along the surface normal. 

\subsection{Spatial distribution and memory}
\label{sec:spatial}

We now discuss two important questions. 
The first concerns the more detailed interpretation of the prefactor 
enhancement. Since this is due to 
confinement of degrees of freedom of the 
molecule, it should be possible to relate it to 
the probability distributions of these degrees of freedom.
The second question concerns the crucial
assumption underlying our calculation of the desorption rate,
that equilibration rates of the adsorbed molecule are
faster than the desorption rate. Equilibration rates are related
to memory times, and we want to use our simulations to
characterise these memory times.

Since H$_2$O contains three atoms, it has nine degrees of freedom.
One of these is bodily vibration along the surface normal, whose
distribution is $y(z)$. The three internal vibrational modes
of the molecule do not undergo large changes on adsorption at low
coverage, so that they do not contribute significantly to the
enhancement of $f$. The enhancement therefore comes from the
remaining degrees of freedom, two of which refer to translation
in the surface plane, and the other three to rotations.
We describe in Appendix~B how the enhancement factor
can be related semi-quantitively to
the non-uniformity of the probability distributions of
these degrees of freedom.

\begin{figure}
\centerline{
\includegraphics[width=4.5in,angle=-90]{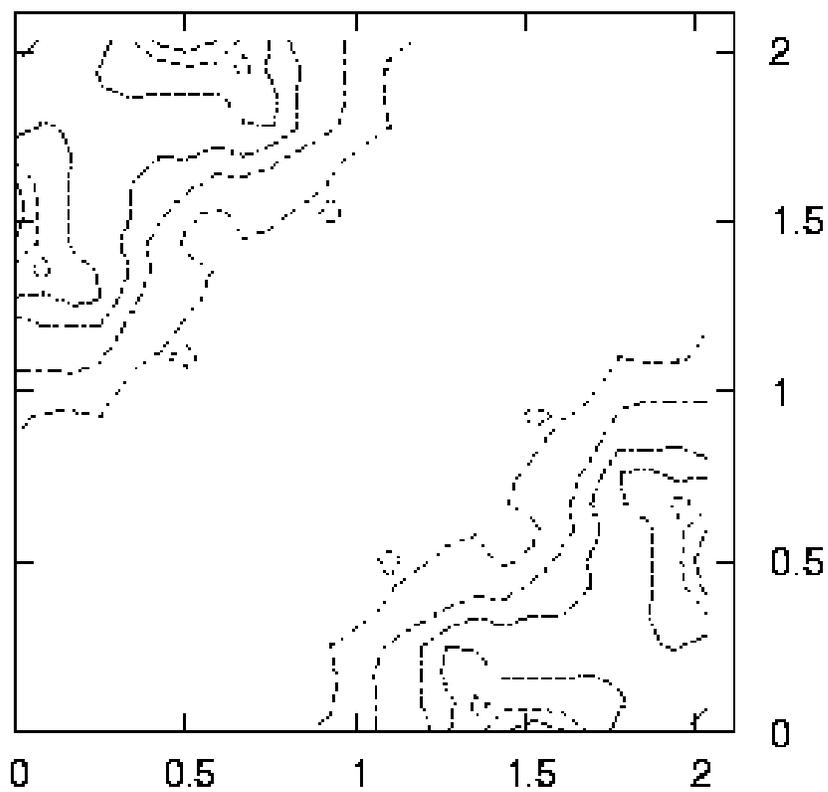}
}
\caption{Contour plot of the spatial probability distribution
of water O atom in the $x$-$y$ plane at $T = 400$~K. Bottom right
and top left corners of plot are Mg sites; bottom left and top right
corners are O sites. Probability density is in arbitary units,
with equal spacing between contours.}
\label{fig:p_trans}
\end{figure}

We examine the reduced freedom of the translational motion by
plotting the probability distribution $p_{\rm tr} ( x, y )$
of the $x$- and $y$-coordinates of the water O atom.
This is done by dividing the surface plane into a square grid,
and accumulating the frequencies with which $x$ and $y$ fall in each
cell of the grid. Since $p_{\rm tr} ( x, y )$ has the
translational symmetry of the crystal surface, we can improve the
statistics by averaging over symmetry related cells.
From our calculated $p_{\rm tr} ( x, y )$ at $T = 400$ 
(Fig.~\ref{fig:p_trans}), we see that the water O spends most of its time close
to surface Mg sites, in line with our finding 
(Sec.~\ref{sec:preliminary_tests}) that
the most stable relaxed configuration of H$_2$O on the surface
has water O coordinated to surface Mg. However, it is also clear that
the most probable O position is not directly above Mg, but is
displaced to positions roughly along a cubic axis in the surface plane.
This agrees with our static calculations on the tilted configuration..
The analysis of Appendix~B shows that we can estimate the
entropy reduction associated with $p_{\rm tr} ( x, y )$ as follows.
We fit the calculated $p_{\rm tr} ( x, y )$
with a superposition of four Gaussians 
$A \exp \left( - \alpha | {\bf r} - {\bf R} |^2 \right)$, centred
at the positions $\delta x ( \pm 1 , 0 )$  and $\delta x ( 0, \pm 1 )$
relative to the
Mg site, using $\alpha$ and $\delta x$ as fitting parameters.
We then normalise the resulting smoothed distribution so that
$\int_{\rm cell} dx \, dy \, p_{\rm tr} ( x, y ) = 1$
where the integral goes over the surface unit cell, whose
area is $A_{\rm cell}$. Denoting
by $c_1$ the constant such that the maximum value of $c_1 p_{\rm tr} ( x, y )$
in the cell is unity, the contribution to the enhancement
factor from translational confinement is estimated as 
$b_1 = A_{\rm cell} / c_1$. At $T = 300$ and 400~K, we find
$b_1 = 3.8$ and 2.9, respectively.

The orientation of the molecule is specified by three angles
$\theta$, $\phi$ and $\psi$. These characterise the orientation
of the molecule relative to a reference orientation, which we
choose to be such that the bisector of the H-O-H bond
points along the outward normal, with the plane of the molecule
lying in the $x$-$z$ plane (the $x$-axis is along one of the cubic
axes of the substrate). From this starting point, we can produce any
other orientation in three steps: (i)~rotate the molecule
about the bisector by angle $\phi$; (ii)~rotate the molecule
by angle $\theta$ about the axis perpendicular to the bisector in the
molecular plane (the
result of this is that the bisector makes an angle $\theta$ with
the surface normal); (iii)~rotate the molecule about the bisector
by angle $\psi$. The angles $( \theta, \phi, \psi )$ are thus
the conventional Euler angles specifying rigid-body rotation of
the molecule.

\begin{figure}
\centerline{
\includegraphics[width=4.5in,angle=-90]{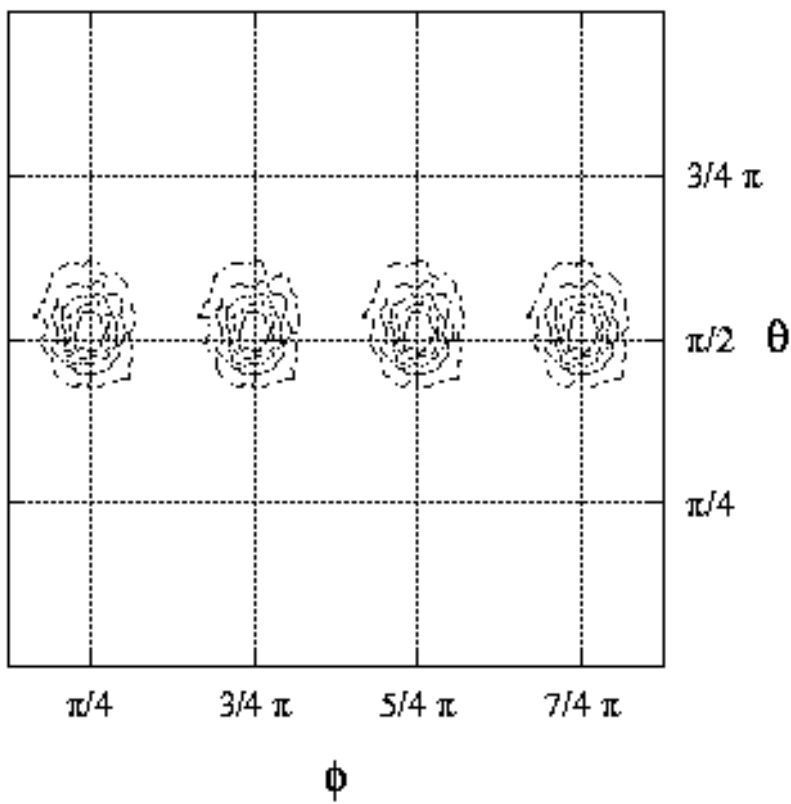}
}
\caption{Contour plot of probability distribution of angles
$\theta$ and $\phi$ specifying orientiation of the bisector of H$_2$O molecule
(see text) at $T = 400$~K. Peaks of the distribution correspond
to the four equivalent orientations in which the bisector is
nearly parallel to the surface ($\theta \simeq \frac{1}{2} \pi$),
and points along one of the diagonal directions ($\phi = \frac{1}{4} \pi$,
$\frac{3}{4} \pi$, $\frac{5}{4} \pi$, and $\frac{7}{4} \pi$).}
\label{fig:p_bis}
\end{figure}

\begin{figure}
\centerline{
\includegraphics[width=4.5in,angle=-90]{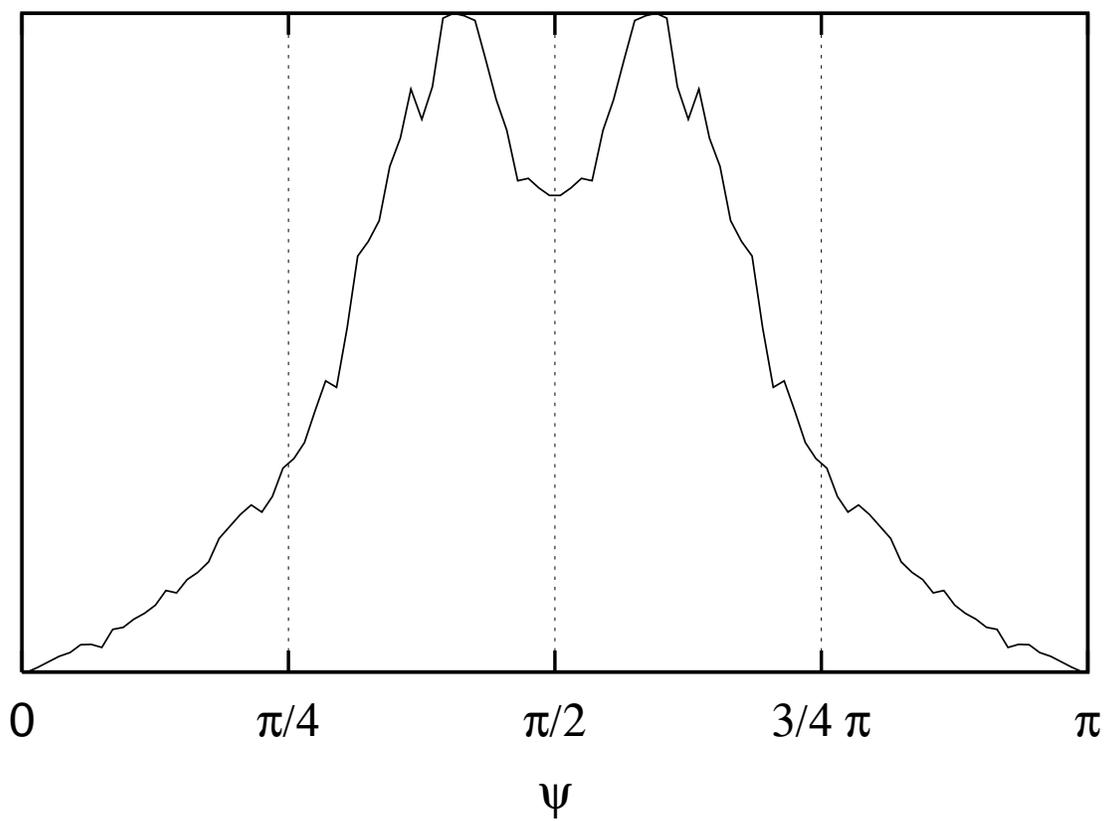}
}
\caption{Contour plot of probability distribution of angle $\psi$
specifying rotation of H$_2$O molecule about its bisector
(see text) at $T = 400$~K. When the molecular bisector is parallel
to the surface, the molecular plane is parallel to the surface when
$\psi = \frac{1}{2} \pi$.}
\label{fig:p_sp}
\end{figure}

The probability distribution $p_{\rm bis} ( \theta, \phi )$ of the 
bisector is computed by dividing the $\theta$ range $( 0, \pi )$ and
the $\phi$ range $( 0, 2 \pi )$ into a uniform grid and accumulating
a two-dimensional histogram. The contour plot of 
$p_{\rm bis} ( \theta, \phi )$ at $T = 400$~K (Fig.~\ref{fig:p_bis}) shows
that the direction of the bisector is always quite close to
the surface plane, and within this plane tends to point along one
of the diagonal directions. (This is exactly the orientation of
the `flat' configuration (Fig.~\ref{fig:adsorbed_config}), in which the two O-H
bonds are directed from a surface Mg site to two of the nearest-neighbour
O sites.) This being so, the angle $\psi$ effectively measures
the tilt of the molecular plane relative to the surface.
The rather broad calculated probability distribution
$p_{\rm sp} ( \psi )$ of the angle $\psi$ (Fig.~\ref{fig:p_sp}), centred on
$\psi = 0$, means that
the molecular plane tilts relative to the surface over a rather
wide range. The double-peaked structure of the distribution
indicates that the untilted configuration is not the most stable,
again in accord with our static results.

The reduction of rotational freedom associated with $( \theta , \phi )$
is estimated by fitting the distribution $p_{\rm bis} ( \theta , \phi )$
to a superposition of Gaussians. As explained in Appendix~B,
the contribution $b_2$ to the enhancement factor due to confinement
of $\theta$ and $\phi$ is estimated by a procedure similar
to the one used for translational confinement. At $T = 300$ and
400~K, we find $b_2 = 19.9$ and 16.0, respectively.
Finally, analysing the confinement of $\psi$ shown in Fig.~\ref{fig:p_sp},
we obtain enhancement contribution $b_3 = 2.8$ and 2.4 at
$T = 300$ and 400~K, respectively. 
Combining all the translational and rotational factors,
we arrive at total enhancement factor of 212 and 111
at $T = 300$ and 400~K, which are 
roughly consistent with what we obtained from
our numerical results for $\gamma$.

Turning now to the question of memory times, we 
want to give evidence
about the dynamics of the two processes that are likely
to have the longest memory times, namely hopping between
different surface sites, and transitions between different
orientations. We have attempted to calculate appropriate
correlation functions, but with only a single molecule in the system,
it is not possible yet to achieve good statistics. The data
we present are therefore only semi-quantitative.

\begin{figure}
\centerline{
\includegraphics[width=4.5in,angle=-90]{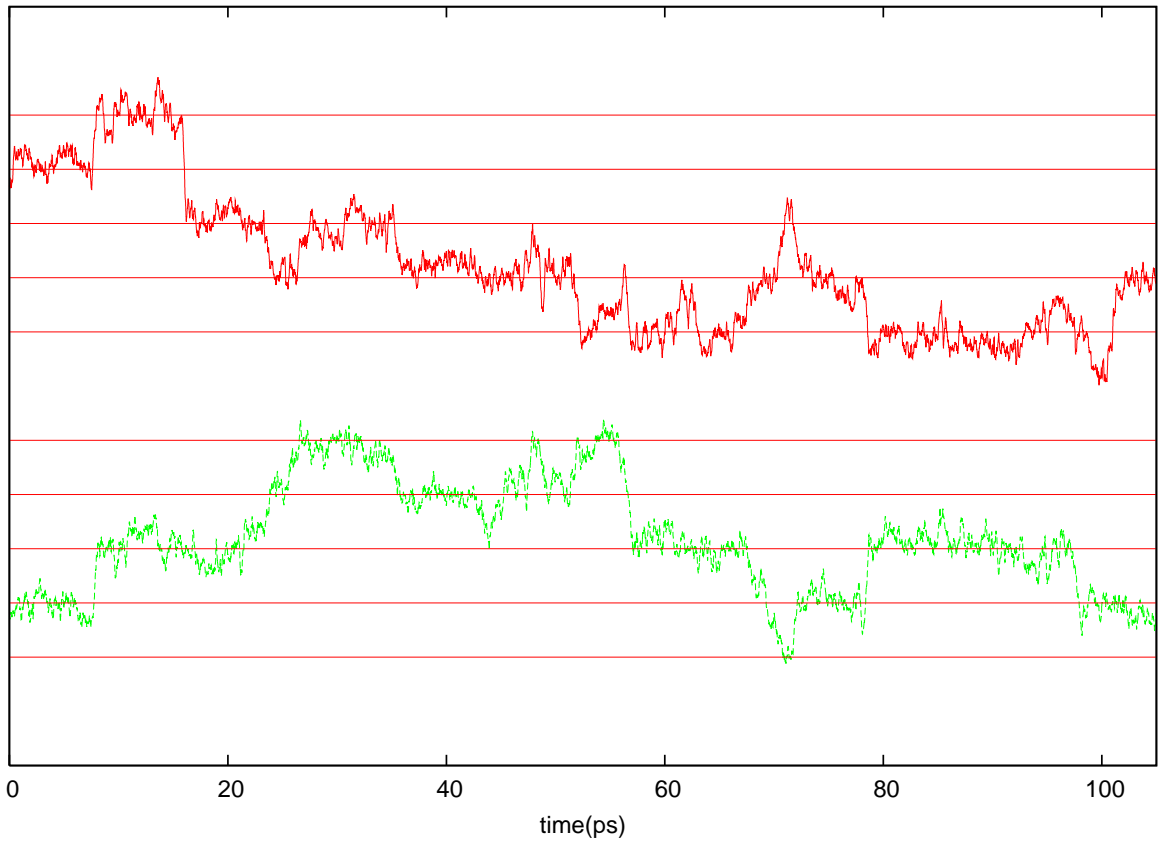}
}
\caption{Time variation of the $x$- and $y$-coordinates of the
water O atom in the course of an m.d. simulation at $T = 400$~K.
Horizontal lines mark $x$- and $y$-coordinates of perfect-lattice
sites, so that spacing between neighbouring lines is
$\frac{1}{2} a_0 = 2.115$~\AA.}
\label{fig:x_y_t}
\end{figure}

To illustrate the translational dynamics, we show in Fig.~\ref{fig:x_y_t} the
time-dependent $x(t)$ and $y(t)$ coordinates of the water O over the
100~ps span of an unconstrained simulation at 400~K. We see
several well-defined intersite jumps in which $x$ and/or $y$
change by the Mg-O nearest-neighbour distance $d = 2.12$~\AA. In most
of these jumps, both $x$ and $y$ change, and this indicates a
diagonal jump between Mg sites, but there are events
in which one coordinate changes by $2 d$, with no change in the other.
By simple counting of jumps,
we estimate a hopping rate of $1.4 \times 10^{11}$~s$^{-1}$. The same
procedure at $T =300$~K gives a hopping rate of $3.8 \times 10^{10}$~s$^{-1}$.
To interprete these results further, we have made nudged elastic band
calculations~\cite{henkelman00} of the 
energy barrier for intersite hopping, and
we find the value 0.13~eV. This is roughly consistent with the ratio
of hopping rates of about 3.7 between 300 and 400~K.
It is clear from this that, in this
temperature region, intersite hopping is very much
more rapid than the desorption rate, and this will become even more
true at lower temperatures.

\begin{figure}
\centerline{
\includegraphics[width=4.5in,angle=-90]{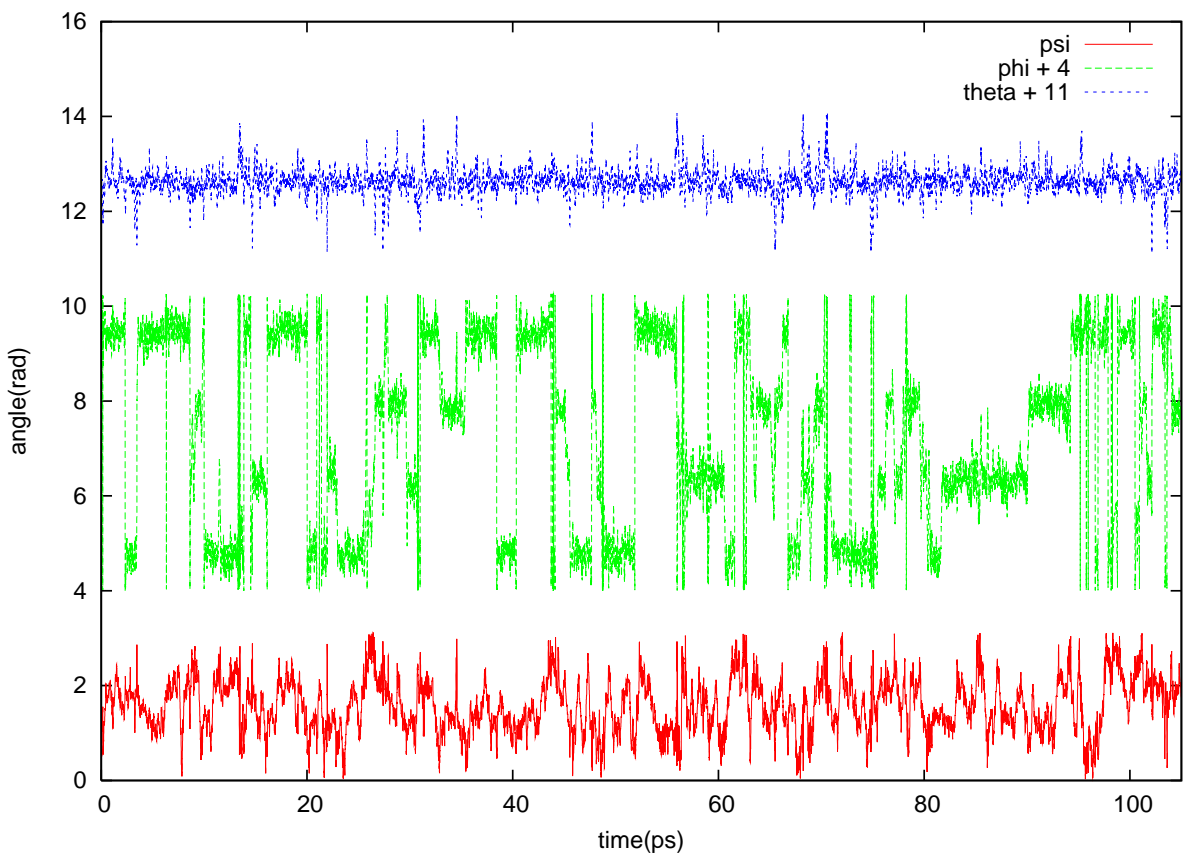}
}
\caption{Time variation of the angles $\theta$ (top), $\phi$ (middle)
and $\psi$ (bottom) specifying orientation of the molecule (see text)
during the course of an m.d. simulation at 400~K.}
\label{fig:theta_phi_psi}
\end{figure}

To display the rotational dynamics, we show 
in Fig.~\ref{fig:theta_phi_psi} plots of
the angles $\theta ( t )$, $\phi ( t )$ and $\psi ( t )$ during the
course of the simulation at 400~K. The dynamics of $\theta$ and
$\psi$ consists of very rapid fluctuations on a time scale of 1~ps or less, over
the range expected from the probability distributions 
of Figs.~\ref{fig:p_bis} and \ref{fig:p_sp}.
The angle $\phi$ has somewhat slower dynamics associated with hopping
between the four equivalent sub-sites around 
each Mg site (Figs \ref{fig:adsorbed_config} and \ref{fig:p_bis}),
but at $T = 400$~K, this hopping rate is $\sim 5 \times 10^{11}$~s$^{-1}$,
which is several times faster than the hopping rate between sites.

The conclusion from this is that all equilibration rates on the
surface appear to be very much faster than the desorption rate
at temperatures of interest.

\section{Discussion}
\label{sec:discussion}

The practical calculations we have presented show the feasibility
of calculating thermodynamic properties of a surface adsorbate
by {\em ab initio} methods, without recourse to a lattice-gas
approximation. We have assumed nothing at all about the adsorption sites
of H$_2$O on MgO~(001), and the {\em ab initio} m.d.
simulations themselves automatically sample the sites and orientiations
that are statistically significant. In the same spirit, we completely
avoid approximations such as the harmonic approximation, and
we fully include the coupling of the molecular and substrate
degrees of freedom. In this sense, there are no statistical-mechanical
approximations whatever, except for the neglect of quantum nuclear
effects, to which we return below. The single uncontrollable
approximation is the DFT exchange-correlation functional.
Furthermore, when calculating the desorption rate, we do not
assume the validity of the Polanyi-Wigner formula, but we
use this formula only as a means of fitting the computed results.
In this way, we obtain an {\em ab initio} value of the
frequency prefactor.

It is well known from both experiments~\cite{paserba01} 
and simulations~\cite{fichthorn02} on a wide range
of systems that frequency prefactors for thermal desorption
often differ, sometimes by many orders of magnitude, from the
value of $10^{13}$~s$^{-1}$ that might naively be expected
if the prefactor is thought of as an ``attempt frequency''.
It is also well known that this is due to the strong reduction
of translational, rotational and conformational entropy
that often occurs when a molecule goes from the gas phase
to the adsorbed state. For H$_2$O on MgO~(001), our calculated
prefactor has the value $f = 2.7 \times 10^{15}$~s$^{-1}$, so that it
is enhanced by a factor of over $100$ above the typical
vibrational frequency of the molecule relative to the surface.
We have seen that an enhancement factor of this general
size is expected from the translational and rotational
probability distribution functions of the molecule.

A crucial assumption behind the Polanyi-Wigner formula is that
the desorption rate depends only on the instantaneous temperature
and coverage, and is not history dependent. A key condition for
this to be true is that the equilibration rate of the molecule
on the surface be faster than the desorption rate. We have attempted
to characterise the typical equilibration times for the isolated
H$_2$O molecule on MgO~(001) by studying the diffusional and
rotational dynamics, and we have found that the condition
is satisfied by a margin of several orders of magnitude in the
temperature region where thermal desorption can be experimentally
observed. 

A direct comparison with experimental TPD data 
for H$_2$O on MgO~(001) is not 
straightforward. There have been several TPD studies 
reported~\cite{stirniman96,xu96}, some of which refer to 
carefully prepared surfaces that appear to be 
relatively free of defects. Desorption at 
sub-monolayer coverage is associated with 
a TPD peak at $\sim 245$~K. In our 
simulated system, the desorption rate at 
this $T$ is $\sim 2.6 \times 10^6$~s$^{-1}$, so that all
molecules would desorb in less than 10~$\mu$s, 
which is many orders of magnitude 
less than the time-scale of a TPD experiment. 
This might suggest that the adsorption 
energy of 0.46~eV given by the PBE functional we 
have used is considerably too low. 
However, the effect of attractive water-water 
interactions may be important even at 
coverages well below the monolayer level. In 
fact, adsorption isotherm measurements 
indicate a critical point in the surface phase 
diagram of H$_2$O on MgO~(001) at 
$T \simeq 210$~K~\cite{ferry96}. The 
effect of water-water interactions on the 
desorption rate clearly needs to be quantified. 
A further complication is that we have 
so far ignored quantum nuclear effects. 
Because of the very high vibrational 
frequencies of the water molecule, it is 
possible that changes of zero-point energy on 
adsorption might shift $E_{\rm ads}$ significantly. 
Even without all these 
effects, it is not clear that we should 
expect good agreement with experiment yet, 
because the calculated $E_{\rm ads}$ depends so 
much on the exchange-correlation 
functional. The frequency prefactor $f$ might 
also depend significantly on exchange-correlation 
functional. If we assume provisionally that our calculated $f$ of 
$2.7 \times 10^{15}$~s$^{-1}$ is essentially correct, 
and we ignore water-water 
interactions, then the experimental TPD peak 
temperature of 245~K would require an 
activation energy $\Delta E = 0.78$~eV, which 
is well above the PBE adsorption energy of 
0.46~eV, though it is still below the LDA value of 0.95~eV. 
It is clear that DFT predictions of 
$E_{\rm ads}$ need to be tested against more 
accurate and reliable methods.

A number of future challenges are suggested by this work. The most obvious
of these is the extension of the calculations to higher
coverages. According to the theory we have presented, the
calculation of the PMF on a chosen molecule, and the integration
of the resulting distribution function $y ( z )$, allows us to
calculate the chemical potential and the desorption rate
at arbitrary temperature and coverage. At the time of writing,
we have performed exploratory {\em ab initio} calculations
of this kind for H$_2$O on MgO~(001) at coverages of 0.25 and 0.5~ML.
However, the problem that emerges is that the memory times
are much longer than for the isolated molecule, so that considerably
longer simulations are needed in order to achieve
acceptable statistical accuracy. This problem of statistical
sampling becomes rapidly worse at low temperatures, and so far we have
achieved stastically accurate results only at $T = 800$~K,
which is well above the region of practical interest.
A second important challenge is that of going beyond
DFT. We have noted that the LDA and GGA forms of exchange-correlation
functional give static adsorption energies differing by roughly a 
factor of two. This means that we must envisage future calculations
of the present kind, but performed by post-DFT techniques. The recent 
proposal of a method for performing quantum Monte Carlo simulations
along an m.d. trajectory~\cite{grossman05} may indicate one
way to do this. Recent successes in applying high-level
quantum chemistry to condensed-matter energetics are
also promising~\cite{manby06}.
A third challenge is that of eliminating the approximation of
classical statistical mechanics for the nuclei. For the case of
H$_2$O on MgO~(001), the errors due to the use of the classical
approximation will not be large, but may be significant if
one wishes to achieve chemical accuracy. We will report in the
second paper of this series on the generalisation 
of the present theory to quantum statistical mechanics for the
nuclei, using path-integral {\em ab initio} simulation.

\section{Conclusions}
\label{sec:conclusions}

In summary, we have shown how {\em ab initio} methods can be used to
calculate the chemical potential of an adsorbate, with full
inclusion of entropy effects. The methods used can in
principle be applied at any coverage. We have shown how the
methods also yield values for the desorption rate and hence
the frequency pre-factor in the Polanyi-Wigner formula.
For the case of H$_2$O on MgO~(001) at low coverage,
this pre-factor is enhanced by at least two orders of magnitude above
the values generally assumed in the past, and we have given a detailed
interpretation of this enhancement in terms of the confinement
of translational and rotational degrees of freedom. The crucial
condition of rapid equilibration necessary for the validity
of the Polanyi-Wigner formula appears to be satisfied by a wide margin
for low-coverage H$_2$O on MgO~(001). Preliminary comparisons
with experimental data suggest that for this system the
adsorption energy given by PBE may significantly too low.

\section*{Appendix~A: Analysis of statistical errors}

We explained in Secs.~\ref{sec:mean_force}
and \ref{sec:results} the importance of monitoring the
statistical errors in the calculation of the mean
force $\langle {\cal F}_z \rangle_z$ and hence of the
PMF and the chemical potential. We summarise here
how we have estimated the errors on the mean force, and how
we have used this to estimate the errors on the
other quantities.

The mean force
is calculated at a set of $z$-values: $z_0 > z_1 , \ldots > z_n$.
At a given $z$-value $z_i$, the estimate ${\cal F}_i$ of the mean
force obtained by averaging over the length of the run differs
from the exact value ${\cal F}_i^{\rm ex}$ that would be
obtained if the sampling were perfect. We denote by 
$\delta {\cal F}_i \equiv {\cal F}_i - {\cal F}_i^{\rm ex}$ the difference
that occurs in a given simulation run. We estimate the standard deviation
$\langle \delta {\cal F}_i^2 \rangle^{1/2}$ by the usual re-blocking method.
In this method the simulation
of total duration $\tau$ is divided into $\nu$ blocks, each of
duration $\tau / \nu$, and we compute an estimate ${\cal F}_i ( l )$
of ${\cal F}_i$ for each block $l$. Then, we compute the
quantity $\sigma_\nu^2 = \nu^{-1} \sum_{l=1}^\nu
\left( {\cal F}_i ( l ) - \bar{\cal F}_i \right)^2$, where
$\bar{\cal F}_i$ is the estimate obtained by averaging over
the entire time $\tau$. If the duration $\tau / \nu$ of each block is
short, the block averages $\bar{\cal F} ( l )$ are strongly
correlated, and $\sigma_\nu$ underestimates the true statistical error.
However, as the duration of the blocks becomes longer than
the correlation time, the ${\cal F}_i ( l )$ become
statistically independent, and $\sigma_\nu$ tends to a plateau value.
The reblocking technique consists of plotting $\sigma_\nu$
against $\tau / \nu$ and taking the standard deviation
$\langle \delta {\cal F}_i^2 \rangle^{1/2}$ to be the plateau
value of $\sigma_\nu$ as $\nu$  is decreased.

We now turn to the statistical error on the PMF, denoting the
value of $\phi ( z )$ at the $i$th $z$-point obtained from
a given set of simulation runs by $\phi_i$, the exact value
by $\phi_i^{\rm ex}$ and the difference $\phi_i - \phi_i^{\rm ex}$
by $\delta \phi_i$. Since the
$\phi_i$ values are obtained by integrating inward from the
largest $z$-value $z_0$ using the trapezoidal rule, we have:
\begin{eqnarray}
\phi_i & = & \phi_0 + \frac{1}{2} \sum_{n=1}^i ( z_{j-1} - z_j )
( {\cal F}_j + {\cal F}_{j-1} ) \nonumber \\
& = & \phi_0 + \frac{1}{2} ( z_0 - z_1 ) {\cal F}_0 +
\frac{1}{2} \sum_{j=1}^{i-1} ( z_{j-1} - z_{j+1} ) {\cal F}_j +
\frac{1}{2} ( z_{i-1} - z_i ) {\cal F}_i \; .
\end{eqnarray}
The errors on ${\cal F}_j$ and ${\cal F}_k$ are statistically
independent for $j \ne k$, so that:
\begin{equation}
\langle \delta \phi_i^2 \rangle = 
\langle \delta \phi_0^2 \rangle + 
\frac{1}{4} \left[ ( z_0 - z_1 )^2 \langle \delta {\cal F}_0^2 \rangle +
\sum_{j=1}^{i-1} ( z_{j-1} - z_{j+1} )^2 \langle \delta {\cal F}_j^2 \rangle +
( z_{i-1} - z_i )^2 \langle \delta {\cal F}_i^2 \rangle \right]  \; .
\label{eqn:delta_phi_sq}
\end{equation}

Now to estimate the statistical error on $\Delta \mu^\dagger$, we recall
(eqn~(\ref{eqn:ECPD_from_mean_force})) that 
$\Delta \mu^\dagger = k_{\rm B} T \ln Y$,
where $Y = d^{-1} \int_{- \infty}^{z_0} dz \, \exp ( - \beta \phi ( z ) )$.
In practice, the integral comes almost entirely from a narrow region 
of $z$ around the value $z_{\rm min}$ where $\phi ( z )$ has its minimum.
But the statistical fluctuations of $\phi_i$ at $z_i$ that 
are near each are almost perfectly correlated, since the
fluctuation $\delta \phi_i$ comes from the accumulation of
fluctuations $\delta {\cal F}_j$ at many points $z_j > z_i$.
So to calculate the fluctuation of $Y$, it is a good approximation
to say that the fluctuations of all $\phi_i$ in the region of
$z_{\rm min}$ are perfectly correlated. This means that
$\delta Y = - \beta \delta \phi_{\rm min} Y$, where
$\delta \phi_{\rm min}$ is the fluctuation of $\delta \phi_i$
for $z_i \simeq z_{\rm min}$. Then since
$\delta \Delta \mu^\dagger = k_{\rm B} T \delta Y / Y$, we finally
obtain the estimate for the standard deviation
$\langle \left( \delta \Delta \mu^\dagger \right)^2 \rangle^{1/2} =
\langle \delta \phi_{\rm min}^2 \rangle^{1/2}$, with
$\langle \delta \phi_{\rm min}^2 \rangle^{1/2}$ calculated according to
eqn~(\ref{eqn:delta_phi_sq}).

%

\section*{Appendix~B: Translational and rotational distributions 
and enhancement of the frequency prefactor}
\label{sec:appendix_b}

We noted in the text that the entropy of the adsorbed molecule is less
than that of the free molecule, because its translational and rotational
freedom is reduced. This entropy reduction is closely related to
the enhancement of the frequency prefactor $f$ above its naively
expected value of $\sim 10^{13}$~s$^{-1}$. We want to make a
semi-quantitative connection between the entropy reduction and the
translational and rotational distributions presented in Sec.~\ref{sec:spatial}.

For this interpretative purpose, we ignore the degrees of freedom
of the substrate, and we assume that the 
$x$-$y$ translational
distribution, the $z$-distribution, 
the $\theta$-$\phi$ distribution and the $\psi$ distribution are all
independent of each other. This is
equivalent to supposing that the distribution over $x$, $y$, $z$,
$\theta$, $\phi$ and $\psi$ is governed by a Boltzmann factor
$\exp ( - \beta V ( x, y, z, \theta, \phi, \psi ) )$, where $V$ is
expressed as:
\begin{equation}
V ( x, y, z, \theta, \phi, \psi ) = - V_0 + \frac{1}{2} \alpha z^2 +
u ( x, y ) + v (\theta, \phi ) + w ( \psi ) \; .
\end{equation}
Here, $- V_0$ represents the energy of the molecule in its most
stable adsorbed configuration, which implies that $u \ge 0$,
$v \ge 0$ and $w \ge 0$. Then the distribution
$y ( z )$ defined in Sec.~\ref{sec:chemical_potential} is given by:
\begin{eqnarray}
\lefteqn{
y ( z ) = C \exp \left[ - \beta \left( - V_0 + \frac{1}{2} \alpha z^2 
\right) \right]
} \\ 
& & \times \int_{\rm cell} dx \, dy \, e^{- \beta u ( x, y )} \;
\int_0^{2 \pi} d \phi \, \int_0^{\pi} d \theta \, \sin \theta 
e^{- \beta v ( \theta, \phi )} \;
\int_0^{2 \pi} d \psi \, e^{- \beta w ( \psi )} 
\end{eqnarray}
for $z$ near the surface, where the $x$-$y$ integration goes over
a single surface unit cell. By definition, $y ( z ) = 1$ far from the
surface, and this fixes the constant $C$:
\begin{equation}
1 = 8 \pi^2 C A_{\rm cell} \; ,
\end{equation}
where $A_{\rm cell}$ is the area of the surface unit cell.
In this approximation, the factor $B$ by which the frequency
prefactor is enhanced is the product of three factors:
$B = b_1 b_2 b_3$, where:
\begin{eqnarray}
b_1 & = & A_{\rm cell} \left/
\int_{\rm cell} dx \, dy \, e^{- \beta u ( x, y )} \right. \nonumber \\
b_2 & = & 4 \pi \left/
\int_0^{2 \pi} d \phi \int_0^{\pi} d \theta \, \sin \theta
e^{- \beta v ( \theta, \phi )} \right. \nonumber \\
b_3 & = & 2 \pi \left/
\int_0^{2 \pi} d \psi \, e^{- \beta w ( \psi )} \right. \; .
\end{eqnarray}
The potential of mean force $\phi ( z )$, defined by 
$y ( z ) = \exp ( - \beta \phi ( z ) )$, is then given by:
\begin{equation}
\phi ( z ) = - V_0 + \frac{1}{2} \alpha z^2 + k_{\rm B} T
\ln ( b_1 b_2 b_3 ) \; ,
\end{equation}
so that the value of $z$ at the minimum of $\phi$ and the
curvature at the minimum remain the same, but the well
becomes less deep with increasing temperature, because of
the factor $k_{\rm B} T \ln ( b_1 b_2 b_3 )$.

We estimate the values of the factors $b_i$ from the
probability distributions as follows. We calculate the
probability distribution $ p_{\rm tr} ( x, y )$ by histogram
accumulation, as described in Sec.~\ref{sec:spatial}, and normalise it
so that $\int_{\rm cell} dx \, dy \, p_{\rm tr} ( x, y ) = 1$.
We then determine the constant $c_1$ such that the maximum
value of $c_1 p_{\rm tr} ( x, y )$ in the
cell is unity ($c_1$ thus has dimensions of area). We then
have $b_1 = A_{\rm cell} / c_1$. Similarly, with
probability distribution $p_{\rm bis} ( \theta , \phi )$ normalised so that
$\int d \phi \, d \theta \, \sin \theta \, p_{\rm bis} ( \theta , \phi ) = 1$,
we find the constant $c_2$ such that the maximum value of
$c_2 p_{\rm bis} ( \theta , \phi )$ is unity. Then $b_2 = 4 \pi / c_2$.
Similarly for $b_3$.

\newpage

\end{document}